\documentclass[aps,pre,twocolumn,superscriptaddress]{revtex4-1}
\usepackage{graphicx,color,colortbl}
\usepackage{epstopdf}
\usepackage{bm}
\usepackage{eucal}
\usepackage{mathrsfs}

\usepackage[dvipsnames]{xcolor}
\usepackage{hyperref}
\definecolor{hypcolor}{named}{BlueViolet}
\hypersetup{colorlinks=true, citecolor=hypcolor, urlcolor=hypcolor, linkcolor=hypcolor}

\usepackage{amsmath}
\usepackage{amsthm}
\usepackage{epsfig}
\usepackage{lipsum}

\begin{document}
\title{Control of Synchronization in two-layer power grids}

 \author{Carl H. Totz}
	\affiliation{Institut f{\"u}r Theoretische Physik, Technische Universit\"at Berlin, Hardenbergstra\ss{}e 36, 10623 Berlin, Germany}
	\author{Simona Olmi}
	\affiliation{Inria Sophia Antipolis M\'{e}diterran\'{e}e Research Centre, 2004 Route des Lucioles, 06902 Valbonne, France}
	\affiliation{CNR - Consiglio Nazionale delle Ricerche - Istituto dei Sistemi Complessi, 50019, Sesto Fiorentino, Italy}
	\author{Eckehard Sch{\"o}ll}
	\email[corresponding author: ]{schoell@physik.tu-berlin.de}
	\affiliation{Institut f{\"u}r Theoretische Physik, Technische Universit\"at Berlin, Hardenbergstra\ss{}e 36, 10623 Berlin, Germany}
	\date{\today}


\begin{abstract}
In this work we suggest to model the dynamics of power grids in terms of a two-layer network, and use the Italian high voltage power grid as a proof-of-principle example.
The first layer in our model represents the power grid consisting of generators and consumers, while the second layer represents a dynamic communication network that serves as a controller of the first layer. In particular, the dynamics of the power grid is modelled by the Kuramoto model with inertia, while the communication layer provides a control signal $P_i^c$ for each generator to improve frequency synchronization within the power grid.
We propose different realizations of the communication layer topology and different ways to calculate the control signal. Then we conduct a systematic survey of the two-layer system against a multitude of different realistic perturbation scenarios, such as disconnecting generators, increasing demand of consumers, or generators with stochastic power output.
When using a control topology that allows all generators to exchange information, we find that a control scheme aimed to minimize the frequency difference between adjacent nodes operates very efficiently even against the worst scenarios with the strongest perturbations.
\end{abstract}


\maketitle



\section{Introduction}

Global warming, the growing world population and power demand, with a subsequent increase in carbon power emissions, have provoked governments and energy utilities to take solid steps
towards the use of renewable energies \cite{UNF15} and their integration within the existing power transmission and distribution systems, thus challenging scientific and technological research towards
the goal of increasing the efficiency and flexibility of the power system \cite{VAC11,JAC11,TUR99,UEC15}. 
The existing power grid was developed using a centralistic approach, therefore we have a few very high-power ac plants operating at 50 or 60 Hz interconnected by ac or dc transmission systems operating at very high voltages (e.g., 400 kV) and many substations, where the high voltage is transformed to the distribution level (e.g., 20 kV). In order to distribute the power in a capillary way, a huge number of distribution lines is present, supplying the loads directly (in the case of high-power loads) or after voltage transformation in the case of residential or low-power industrial loads (e.g., 400 V in Europe).
Recently, renewable energy generators, which produce a few kilowatts in the case of residential photovoltaic systems, up to some megawatts in the case of large photovoltaic
and wind generators, have become widely dispersed around the world, thus transforming the present power system into a large-scale distributed generation system incorporating
thousands of generators, characterized by different technologies, voltage, current, and power levels, as well as topologies \cite{Kanchev11, Ramachandran11}. 
Hence, their integration with the existing network is fundamentally changing the whole electrical power system \cite{Yang11,JAC11}: the drawback of renewable energy power plants 
is that their output is subject to environmental fluctuations outside of human control, i.e., clouds blocking the sun or lack of wind, and these fluctuations emerge on all timescales displaying 
non-Gaussian behaviour \cite{MIL13,HEI10}. In addition, these issues are further complicated by the aging infrastructure of the existing power grid, which already cause problems to utilities and customers, providing low power quality at increasing cost.
In particular the power grid infrastructure is very critical and contains a large number of interconnected components: generators, power transformers, and distribution feeders that are geographically
spread. Moreover, its increasing complexity and geographical spread, and the side effects caused by the high penetration of renewable, stochastically fluctuating energy generators make it very vulnerable, both from the point of view of required sophisticated security mechanisms \cite{Morante2006} and from the point of view of dynamic stability, since 
renewable sources are usually employed by microgrids in isolated modes to maintain their capability of connecting and disconnecting from the grid \cite{Balaguer2011}.
Due to the design of the current power grid as a centralized system where the electric power flows unidirectional through transmission and distribution lines from power plants to the 
customer, the control is concentrated in central locations and only partially in substations, while remote ends, like loads, are almost or totally passive. 
Therefore it is necessary to design new systems that provide more effective and widely distributed intelligent control embedded in local electricity production, two-way electricity and information flows, thus achieving flexible, efficient, economic, and secure power delivery \cite{Liserre2010}.

The new approach, widely known as Smart Grid \cite{Santacana2010}, requires both a complex two-way communication infra\-structure, sustaining power flow between intelligent components, and sophisticated computing and information technologies, as well as business applications. The new approach will include grid energy storage, needed for load balancing and for overcoming energy fluctuations caused by the intrinsic nature of renewable energy sources, in addition to preventing widespread power grid cascading failures \cite{Calderaro2011, Bakken2011}.
In particular, control is needed in power networks in order to assure stability and to avoid power breakdowns or cascading failures: one of the most important control goals is the preservation of synchronization within the whole power grid. Control mechanisms able to preserve synchronization are ordered by their time scale on which they act: 
the first second of the disturbance is mainly uncontrolled, and in this case a power plant will unexpectedly shut down with a subsequent shortage of power in the system, energy is drawn from the spinning reserve of the generators. Within the next seconds, the primary control sets on to stabilize the frequency and to prevent a large drop. Finally, to restore the frequency back to its nominal value of 50 (or 60) Hertz, secondary control is necessary.  In many recent studies on power system dynamics and stability, the effects of control are 
completely neglected or only primary control is considered \cite{Dorfler13, SCH18c, ROH12, SCH15, SCH16, WANG16}. This control becomes less feasible if the percentage of renewable power plants increases, due to their reduced inertia \cite{ULB14,DOH10}. Few studies are devoted to secondary control \cite{Weitenberg18, Tchuisseu18, Simpson12} and to time-delayed feedback control \cite{Okuno2006, Dongmo17, TAH19}. 

The aim of this work is to investigate the controllability of power networks subject to different kinds of perturbations and to develop novel control concepts considering the communication infrastructure present in the smart grid.
Few works have included the communication layer into the synchronization of power networks. Even though the communication infrastructure plays an important role in control and synchronization, preliminary 
works \cite{Li11, Wei12} assume trivial networks, without disconnected nodes, which, however, is of great importance in stabilizing smart grids, due to the necessity of synchronizing grids with isolated generators, microgrids, or even coupled microgrids that can be connected or disconnected to the main grid at any time. 
Moreover, the inclusion of a communication infrastructure has added new challenges in control and stability \cite{Baillieul2007}, where communication constraints emerge, e.g., time­-delays, packet losses, sampling and data rate, among others, but, up to now, attention has focussed on sampling problems in order to assure that synchronization is independent on the sampling period \cite{Giraldo:CDC:2013}. 

In this paper we consider a two-layer network in a full dynamic description. It consists of a power grid layer and a communication layer, which provides the control for the power grid.
Each layer is governed by its own dynamics, which is dependent upon the state of the other layer. In particular the physical topology that relates the interconnection of distributed generators and loads is described by coupled Kuramoto phase oscillators with inertia, closely related to the swing equations \cite{Filatrella:EPJB:61}, while 
the communication topology, which describes the information flow of the power system control measurements, depends on the information of the neighbors of each node \cite{Giraldo:CDC:2013}.
Starting from the ideal synchronized state, we investigate the effect of multiple different perturbations to which the system is subject, modelling real threats to synchronization of the network, e.g., failure of nodes, increased consumer demand, power plants with stochastically fluctuating output. For each perturbation different setups of the communication layer are tested to find an effective control strategy that successfully preserves frequency synchronization against all applied perturbations. 
As a proof of concept the Italian high voltage power grid has been considered. The same two-layer topology has been already investigated in \cite{BUL10} to understand 
how localized events can present a severe danger to the stability of the whole power grid, by causing a cascade of failures, but without considering the dynamics of the control nodes. 
Here the focus of our investigation is on the interdependence of the communication network and the power grid: Random failure of a power plant causes the malfunction of connected elements in the communication layer. Communication nodes isolated due to the failure become inert, causing generators connected to them to shut down as well as eventually leading to a far-reaching blackout.
In short, our proposed control techniques preserve synchronization for different perturbations, thus demonstrating the powerful perspectives of our control approach which considers synchronization of power systems based on the coupled dynamics of the smart grid architecture and the communication infrastructure.

\section{Results}

\subsection{Typical perturbation patterns}
\begin{figure*}[ht]
	\centering
	\includegraphics[width=0.33\linewidth]{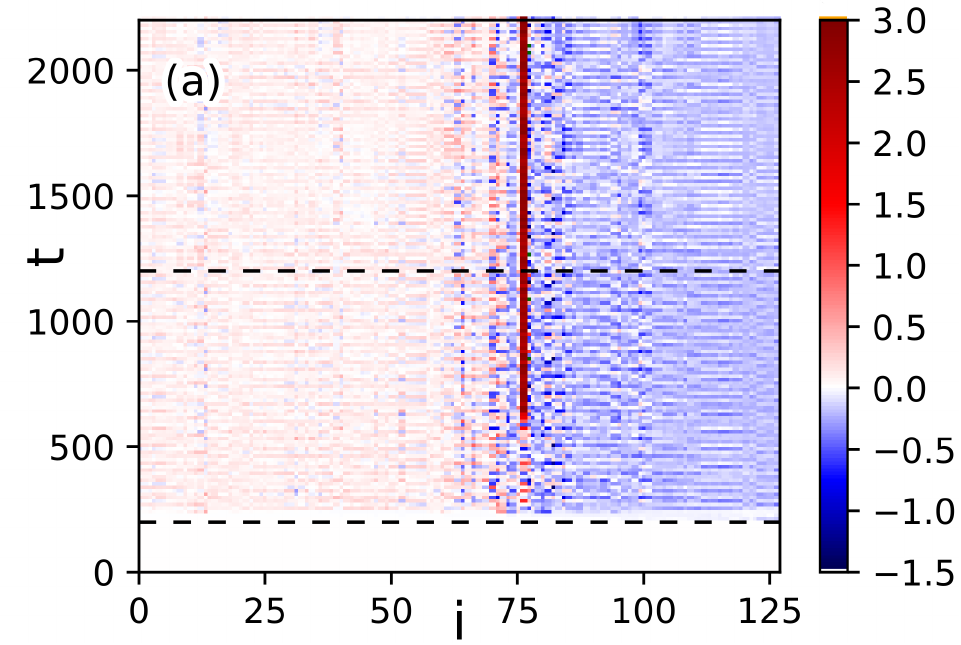}
	\includegraphics[width=0.32\linewidth]{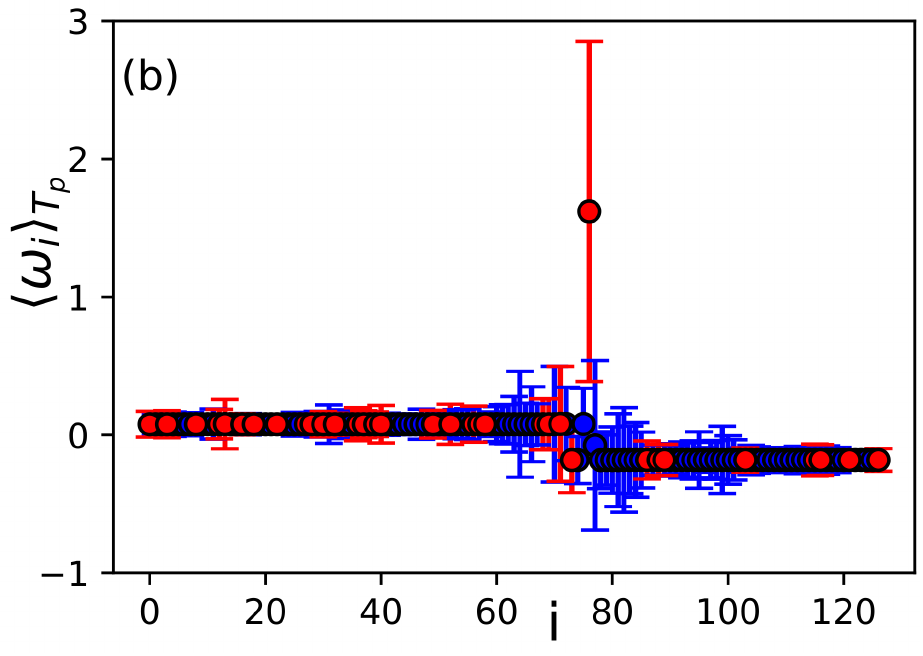}
	\includegraphics[width=0.33\linewidth]{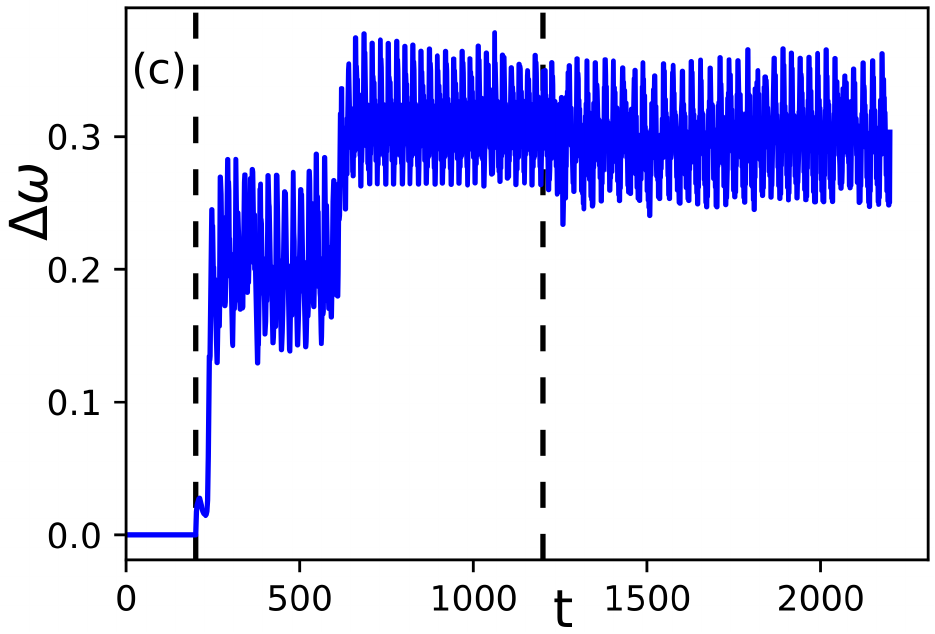}
	\caption{
		Typical desynchronization pattern emerging due to the application of perturbations without control. Here the demand of load $i=120$ is increased to $\Omega_{pert}=-3$ for the duration of the perturbation $T_p=1000$ from $t=200$ to $t=1200$, marked by dashed lines.
		(a) Space-time plot of the whole Italian power grid. Colour indicates the frequencies of the nodes, whose values are reported in the corresponding colour scale on the right. The nominal intrinsic frequency is normalized to zero.
		(b) Mean frequency of the nodes during the time of perturbation $\langle\omega_i\rangle_{T_p}$. Blue (red) nodes identify generators (loads).
		(c) Standard deviation of the ensemble frequency $\Delta\omega$ vs time.
		Coupling strength $K=6.5$,
		inertial mass $m=10$, 
		bimodal frequency (power) distribution $\Omega_{load}=-1$ and $\Omega_{gen}=\frac{93}{34}$.
		Integration time step $\Delta t=0.002$. 
	}
	\label{fig:perturbation_pattern}
\end{figure*}

Applying a sufficiently strong perturbation to the synchronized state of the network causes a loss of full frequency synchronization.
Due to the elongated structure of the Italian power grid, the main pattern emerging in this case is the desynchronization between the northern and southern parts.
A typical example of this event is illustrated in Fig.\thinspace\ref{fig:perturbation_pattern}\thinspace (a), where the perturbation pattern following increasing demand in a single node is shown. The unperturbed nominal intrinsic frequency is normalized to zero.
In particular, as a consequence of the applied perturbation, nodes in the northern part of the network (nodes approximately with index $i\leq70$) adopt a slightly higher positive frequency, 
while frequencies related to the nodes in the southern part shift to negative values.
This is due to the unbalanced distribution of generators in the network: The southern part of the grid possesses a higher ratio of loads 
(11 generators and 46 loads), while the northern part contains a higher ratio of generators (23 generators and 47 loads).
Therefore, when frequency synchronization between the two parts of the power grid is lost, they adopt the mean frequency of their subgroup ($\overline{\dot{\vartheta}}_{north}\approx0.23$ and $\overline{\dot{\vartheta}}_{north}\approx-0.28$ respectively).
The boundary between the southern and the northern part is identifiable when looking at the node frequencies, averaged over the perturbation time $T_p$.
Slight fluctuations are present across the whole network, as indicated by the error bars in Fig.\thinspace\ref{fig:perturbation_pattern}\thinspace (b), while
fluctuations become stronger near the boundary of the two parts of the grid, i.e., for $i\in\left[60,\,90\right]$. 
\\
Moreover, as a consequence of the increasing demand (single node perturbation), 
we observe a macroscopic reaction of the network, whose standard deviation of the frequency increases drastically and oscillates in time (see Fig.\thinspace\ref{fig:perturbation_pattern}\thinspace (c)).
\\
Finally, a single node perturbation can cause the destabilization of a distant node, e.g., fluctuations often cause generators near the boundary between north and south to desynchronize.
Especially susceptible are the generators $i=71,\,76$, which are located as dead-ends along the central connection between the northern and southern part. 
An example of generator $i=76$ being desynchronized, as a result of perturbing a remote load, is illustrated in Fig.\thinspace\ref{fig:perturbation_pattern}\thinspace (b).
\\
In summary, perturbation patterns are multiple and diversified, as will be shown in detail and more systematically in the following sections, where different applied perturbations are considered. 
In particular detailed space-time plots showing all the perturbation patterns are shown in the Supplemental Material \cite{suppl}.
Therefore it is necessary to design a proper control scheme which is able to cope with different scenarios, via a rapid exchange of information, and this is achieved by the communication layer which we will introduce in order to effect control.

\subsection{Disconnecting nodes}
\begin{figure*}[ht]
		\includegraphics[width=0.32\linewidth]{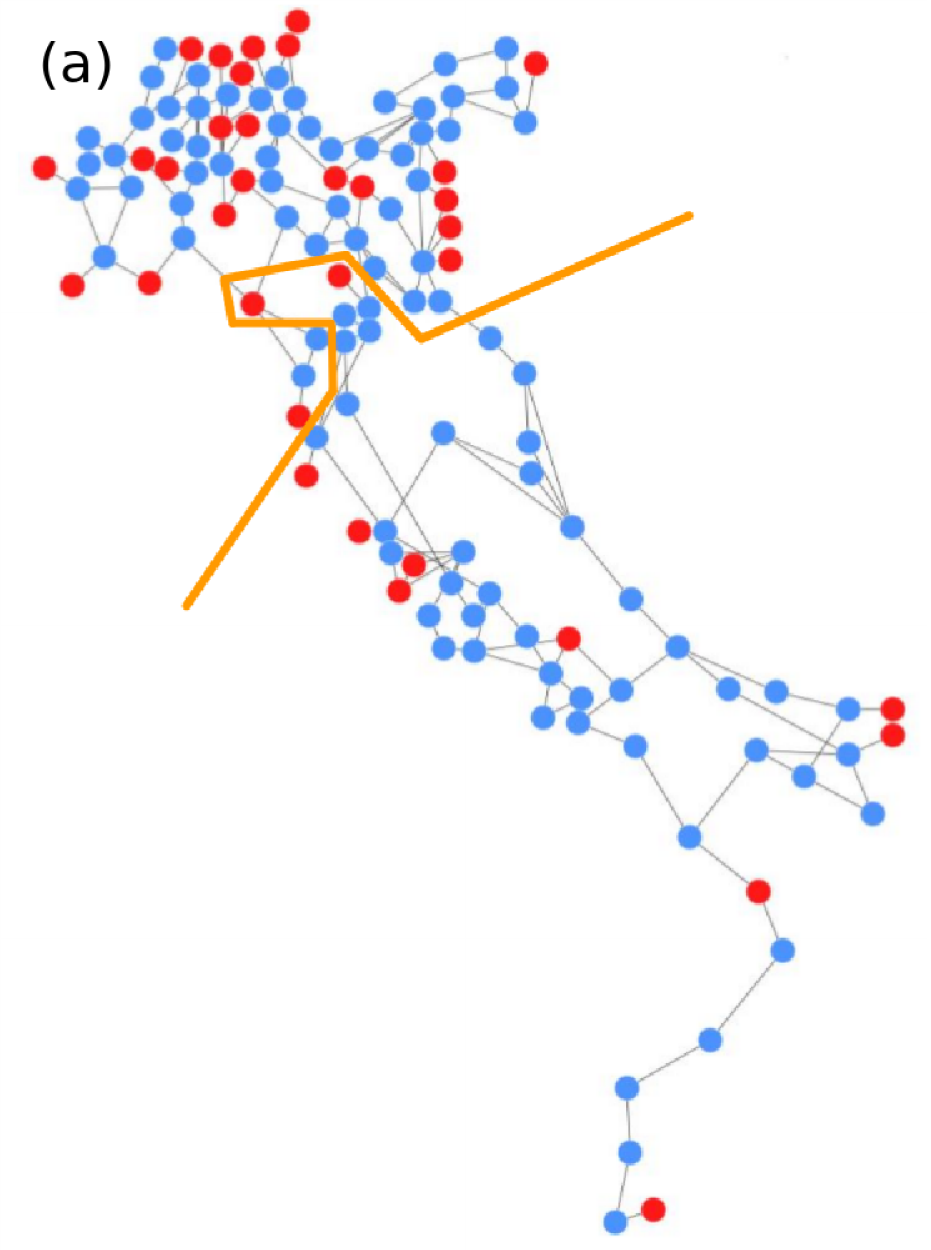}
		\includegraphics[width=0.32\linewidth]{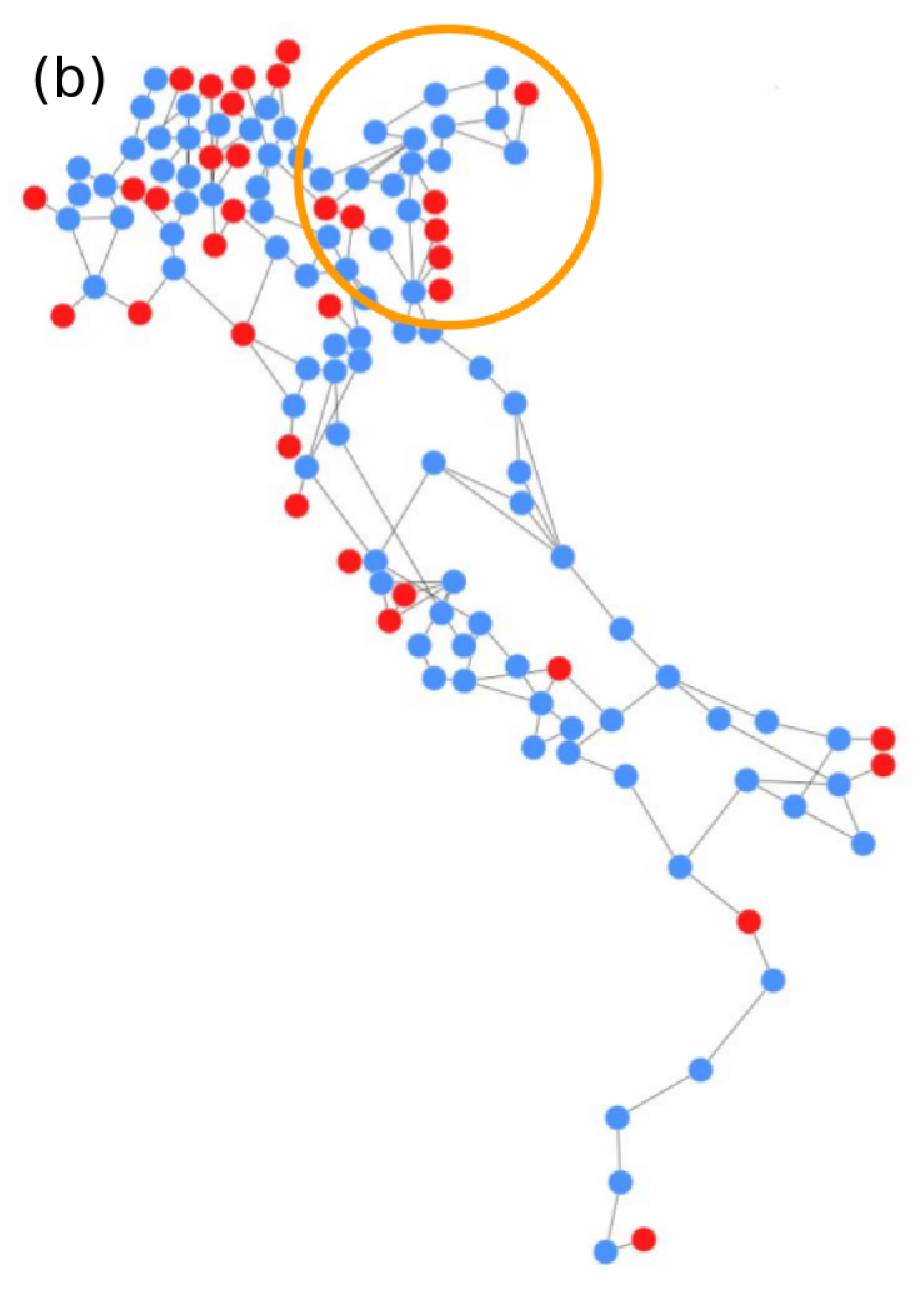}
		\includegraphics[width=0.32\linewidth]{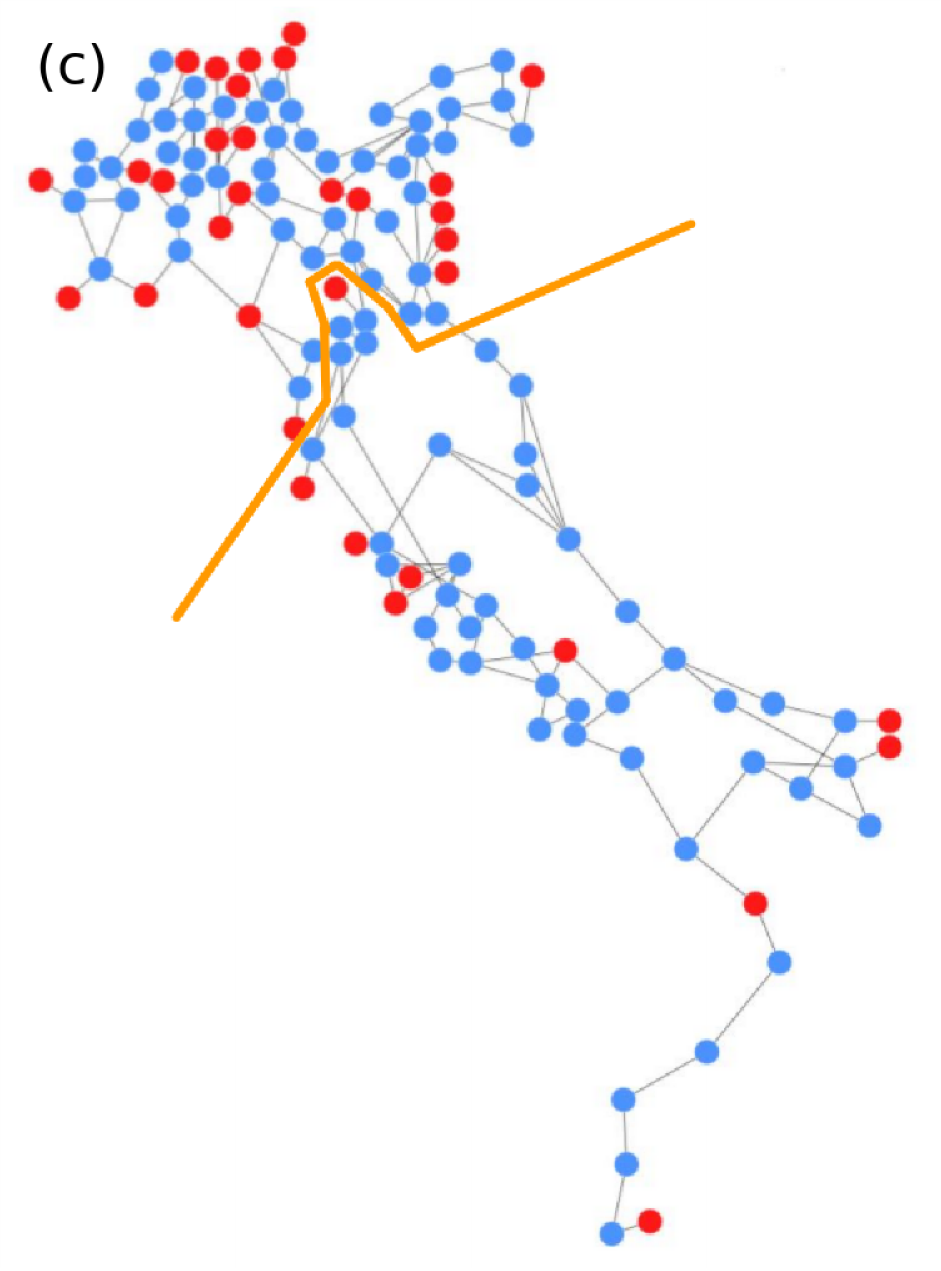}
	\caption{Map of the Italian high voltage power grid with 34 generators and 93 loads. Generators are marked by red dots, while blue dots indicate loads. For the numbering of the nodes refer to the Supplemental Material \cite{suppl}. Frequency synchronization of the network is vulnerable to disruption of nodes depending on their location and the type of perturbation: 
	(a) Disconnection of any generator below the orange line is critical to the frequency synchronization of the whole power grid.
	(b) Applying intermittent noise to any generator outside the orange circle is critical to frequency synchronization.
	(c) Instantaneous increase in demand of any load below the orange line is critical to frequency synchronization.
	}
	\label{fig:VulnMaps}
\end{figure*}

First, we systematically investigate the effect of disconnecting a generator on the synchronization of the network by targeting each generator in the network individually. The generator $i$ is disconnected from the power grid and the communication layer as defined in the Methods Section IV.D 
for the duration of $T_p=1000$ time units.
The perturbation affects the grid depending on the topology: it turns out that nodes in the southern part of the network are particularly vulnerable to selected disconnection, while
the power provided by the generators in the northern part can easily be replaced by the power of the others when disconnected.
If no control is applied to the generators, the network loses synchronization, as soon as any generator in the southern part of the network is targeted by this perturbation. Fig.\thinspace\ref{fig:VulnMaps} shows the Italian high voltage power grid with 34 generators and 93 loads. In particular in Fig.\thinspace\ref{fig:VulnMaps}\thinspace (a) the boundary is indicated at which desynchronization is observed: only below the orange line, the targeted disconnection results in loss of synchronization. Vulnerability to other kinds of perturbations in shown in (b) (intermittent noise applied to a generator) and (c) (instantaneous increase in demand of a load).

\begin{figure}[ht]
		\includegraphics[width=1\linewidth]{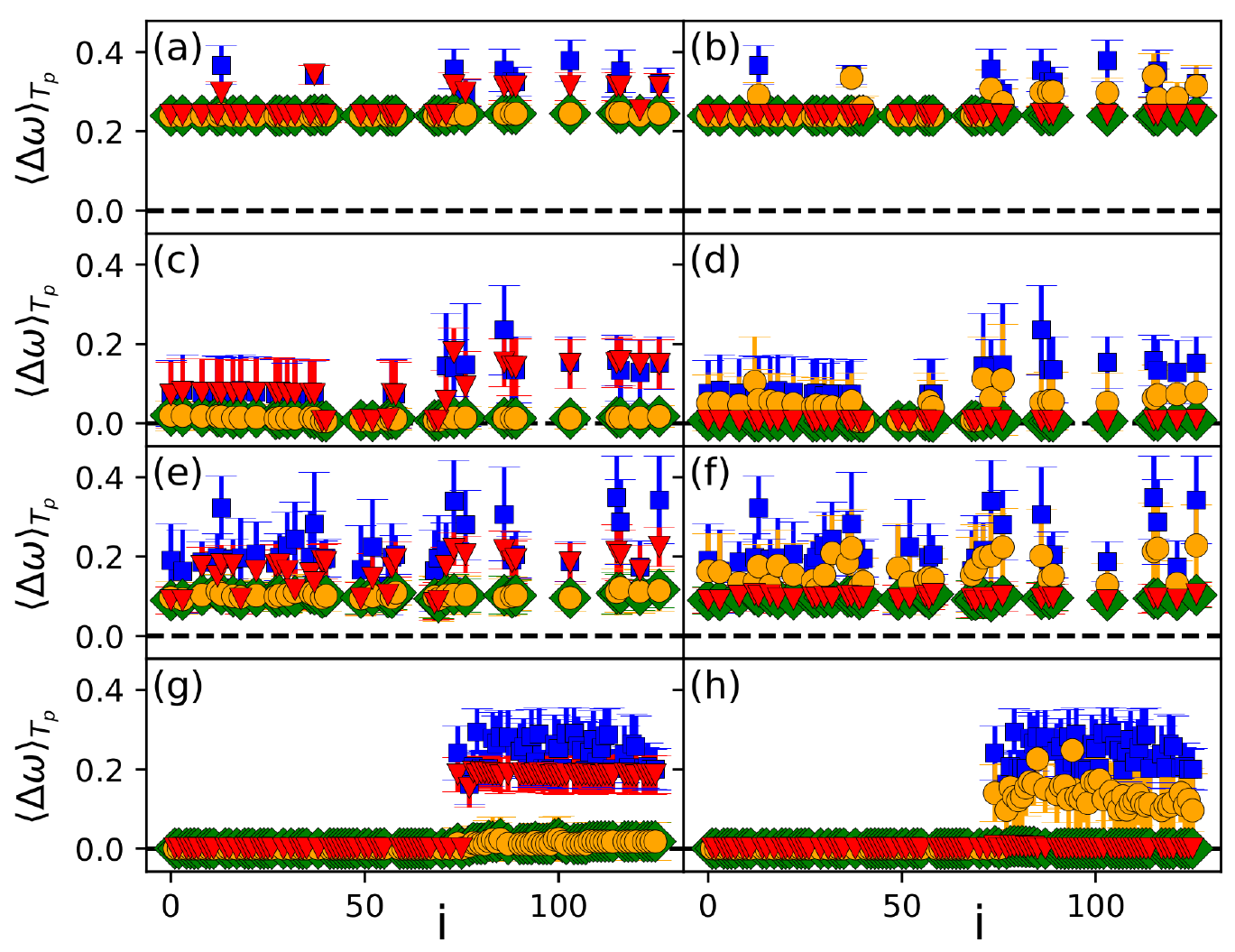}
	\caption{Control of frequency synchronization for different perturbations targeting a single node: Mean standard deviation of frequency during the time of perturbation $\langle\Delta\omega\rangle_{T_p}$ vs the index $i$ of the targeted node. 
	The symbols indicate different control schemes. Blue squares: no control. Red triangles: \textit{difference control}. Yellow circles: \textit{direct control}. Green diamonds: \textit{combined control}.
	Control strength: $G=0.04$.
	The two columns show different control layer topologies: The left column shows the simple setup ($c_{ij}^{loc}$) of the control layer, where generators possess the same connections as in the power grid layer, and the right column shows a control layer topology ($c_{ij}^{ext}$) where additional links between all generators are present.
	The four rows indicate different types of perturbations:
	(a),(b): Disconnecting nodes using (a) $c_{ij}^{loc}$, (b) $c_{ij}^{ext}$.
	(c),(d): Intermittent noise with parameters $\mu=\Omega_{gen}$, $\sigma_x=\frac{1}{3}$, $I=2$, $h(f)=f^{-\frac{5}{3}}$, $g=0.5$, $z_0=2$, $\gamma=1$ using (c) $c_{ij}^{loc}$, (d) $c_{ij}^{ext}$.
	(e),(f): White noise with $\sqrt{2D}=5.0$ using (e) $c_{ij}^{loc}$, (f) $c_{ij}^{ext}$.
	(g),(h): Increasing load demand to $\Omega_{pert}=-3$ using (g) $c_{ij}^{loc}$, (h) $c_{ij}^{ext}$.
	}
	\label{fig:CMode_Comp_single}
\end{figure}

If any generator in the southern, peninsular part is disconnected, it causes not only the southern part, but also the northern,  continental part of the network to desynchronize as a natural consequence.
If any of the northern generators is targeted, this generator returns to its natural frequency $\Omega_{gen}$ due to inertia, as it is no longer connected to any other node in the network, while the rest of the network remains generally unaffected by the disconnection.
 %
The only exception to this is given by generator $i=37$, close to the northern border of the power grid, at the border with Switzerland. When this generator is disconnected, the perturbation causes node $i=36$ to become isolated from the rest of the power grid since it is directly connected only to node $i=37$, thus the entire dead tree desynchronizes.
%
Another singular example is generator $i=121$. The disconnection of this node causes Sicily to become isolated from the main part of the power grid. During the isolation both the main part of the power grid and the Sicilian subset remain frequency synchronized within themselves, but after the connection is restored, the isolated part is unable to resynchronize with the main part of the grid without control. 
\\
%
%
In the following we show that control of the generators during the perturbation is able to counteract the desynchronization, and we analyze which of the control schemes are more effective than others. We study three different control functions as detailed in the Methods Section: (i) difference control, (ii) direct control, (iii) combined control of both.
When the topology of the controlling communication layer is chosen identical to the power grid layer (Fig.\thinspace\ref{fig:CMode_Comp_single}\thinspace (a)), we observe that \textit{difference control} turns out to be less effective in counteracting the perturbation than the other two control schemes, \textit{direct} and \textit{combined control}, which offer the same effectiveness.
This occurs due to the properties of the control schemes. For example, \textit{difference control} only preserves frequency synchronization locally, but due to a lack of connections between the northern and southern part of the power grid, this scheme is only able to stabilize both parts of the network separately, while their mutual frequency desynchronization remains.
When additional links between the northern and southern part of the network are introduced in the control layer (Fig.\thinspace\ref{fig:CMode_Comp_single}\thinspace (b)), \textit{difference control} becomes able to restore full frequency synchronization within the network, albeit with a nonzero offset due to the missing generator, as discussed below.
\\
For \textit{direct control} the behavior is the opposite. 
When additional links are present in the communication layer between the generators in the northern and southern part (right column), this scheme becomes inert, as the mean frequency within the vicinity of each generator is identical to the nominal frequency of the network.
Without additional links, however, this control scheme is able to restore the nominal frequency of the network in the vicinity of each generator and thus in the whole network (left column).
The \textit{combined control} always operates closely to the more effective of the two control schemes (\textit{difference control}, \textit{direct control}).
\\ 
Note that the presence of a nonzero offset in the mean standard frequency deviation, when $\langle\Delta\omega\rangle_{T_p}$ is plotted as a function of the targeted node, has been artificially introduced by the performed numerical procedure, since the standard deviation is calculated considering all nodes in the network, including the disconnected generator. Because it is disconnected, the frequency of this generator returns to its natural frequency $\Omega_{gen}$. As a consequence the ensemble standard deviation of frequencies of the power grid is non-zero, even if the rest of the network is fully frequency-synchronized.
\subsection{Applying intermittent noise to generators}
A characteristic feature of renewable energy sources are power fluctuations due to fluctuating wind and solar irradiation (clouds).
This requires novel design concepts and theoretical investigations into smart storage control strategies to balance feed-in variations and mitigate power quality problems induced by stochastic fluctuations. A particular challenge for stable power grid operation is given by wind- and solar-induced fluctuations, which follow characteristic non-Gaussian statistics over a broad band of time scales from seasonal and diurnal imbalances down to short-term fluctuations on the scale of seconds and sub-seconds \cite{Anvari2016}. The turbulent character of wind feed-in, and in particular its intermittency, is directly transferred into frequency and voltage fluctuations, as shown in \cite{Schmietendorf:EPJB:90}, where the main characteristics of real wind feed-in were captured by generating intermittent time series on the basis of a Langevin-type model and imposing a realistic power spectrum.
In the following we apply intermittent noise, generated as in \cite{Schmietendorf:EPJB:90}, and we systematically investigate the effect of applying noise to each individual generator.
For the detailed description of the perturbation procedure see Sec.~\ref{perturbations}. 
In particular, for the stochastic term $\Omega_N(t)$ 
we use intermittent noise with high intermittency strength $I=2$. 
The value of the intermittent noise $x(t)$ is restricted to $-1\leq x \leq 1$ and the penetration parameter $\mu=\Omega_{gen}$ is chosen identical to the power output of each generator, to prevent any generator from acting as a load and giving a maximum power feed-in to the network.
\\
When the perturbation is applied systematically to each individual generator in the network in the absence of control, we observe that the whole network always loses synchronization, irrespectively of the targeted node (Fig.\thinspace \ref{fig:CMode_Comp_single}\thinspace (c), (d)).  Due to the perturbation, the frequencies of the southern part of the network usually start fluctuating with slightly negative frequencies, while the frequencies of the northern part fluctuate correspondingly with slightly positive frequencies. This behavior emerges during the perturbation and remains even after the perturbation ends. However generators do not usually desynchronize from the rest of the power grid, apart from a few cases (i.e., nodes $i$=86, 115), which correspond to dead ends. 
Moreover fluctuations affect the network with a different timing, depending on the topological location: if a node belonging to the southern part of the network is targeted, the network loses synchronization almost immediately; if a node belonging to the north-western part of the power grid is targeted, the power grid eventually loses frequency synchronization after some time $t\sim T_p/2$.
Only the generators in the north-eastern part of the power grid ($i=39,\,40,\,49,\,52,\,56,\,68,\,69$) remain resilient as highlighted in Fig.\thinspace \ref{fig:VulnMaps}\thinspace (b).

In Fig.\thinspace \ref{fig:CMode_Comp_single}\thinspace (c),\thinspace (d) the ability of the different control schemes to preserve frequency synchronization in the presence of the perturbation is illustrated.
We observe similar behavior as in the case of disconnected single generators: \textit{difference control} is only effective in counteracting the perturbation if additional links between the generators are present in the communication layer, while \textit{direct control} is only effective in the absence of additional connections.
In particular, when applying \textit{difference control} to a sparse control network (i.e. generators in the control layer are connected just to their direct neighbors),
the underlying power grid is unable to recover full frequency synchronization after the perturbation ends, since the frequency shift between the northern and the southern parts remains unchanged.
If the generators in the control layer are globally connected, full frequency synchronization is always achieved after the end of the perturbation with \textit{difference control}, except if node $i=73$ is perturbed: in this case short-living fluctuations emerge, on a time $t\sim 150$; they possess an intensity of $\Delta \omega \sim 0.2$. 
On the other hand, the application of \textit{direct control} to a local control layer topology allows for the achievement of full frequency synchronization at the end of the perturbation. Moreover full frequency synchronization is mostly retained even during the perturbation, irrespectively of short-living fluctuations comprising the whole network that emerge for a time $t\sim 100$. When \textit{direct control} is applied to the extended control layer topology, the frequency synchronization is not retained during the perturbation and a frequency shift between the northern and southern parts emerges.
Finally, \textit{combined control} is as effective as the more efficient of its two components depending on topology, as the ineffective component is mostly inactive.
\subsection{Applying Gaussian white noise to generators}
In the previous section we have investigated the impact of short-term wind fluctuations on the basis of a Kuramoto-like power grid model, considering stability in terms of frequency desynchronization. In this section we compare intermittent feed-in  with Gaussian white noise, and investigate the impact of uncorrelated noise upon the generators. In particular each generator is perturbed by applying Gaussian white noise with intensity $\sqrt{2D}=5.0$ (for the implementation of noise in the dynamical equations, see Sec.~\ref{perturbations}). 
\\
If white noise is applied to each generator individually we observe a general vulnerability of the network, independently of the perturbed node (see Fig.\thinspace\ref{fig:CMode_Comp_single}\thinspace (e),\thinspace (f)).  Without control, the network is usually unable to recover after the perturbation, and frequency desynchronization between the northern and southern part of the power grid occurs. Generally, the desynchronization is stronger than for intermittent noise (c, d). Almost all generators are vulnerable except for a few resilient ones (i.e., $i=3,\,18,\,30,\,36,\,49,\,56,\,68,\,69$), which are located in the northern part of the grid, though not located closely together. 
If one of these resilient nodes is targeted, the perturbation only causes long-time frequency fluctuations in the neighborhood of the affected generator, while the remaining part of the network does not lose full frequency synchronization.  
\\
If we implement the control strategy, as shown in Fig.\thinspace\ref{fig:CMode_Comp_single}\thinspace (e),\thinspace (f), we observe as before that without additional links between the generators, \textit{difference control} is unable to reliably restore synchronization within the whole power grid (e).
When links between all generators are added (f), the control scheme becomes very effective in counteracting this perturbation.
Likewise \textit{direct control} is only able to effectively ensure frequency synchronization within the whole power grid in the absence of additional links (e).

\subsection{Increasing demand of loads}
Perturbations can be applied not only to generators but also to loads. In particular, we investigate the effect of instantaneously increasing the demand of a single load upon the frequency synchronization of the network. We apply this perturbation as defined in Sec.~\ref{perturbations} 
by increasing the demand of the individual load $i$ by a factor of three to $\Omega_{pert}=-3.0$, for the duration of the perturbation, in order to represent sudden changes in consumer behavior due to singular events.
%
\\
The impact of increasing the demand of a single load strongly depends upon the node location and we observe a distinct topological separation between vulnerable and resilient nodes.
If any node in the northern part of the network ($i\leq 72$) is perturbed, the system remains unaffected: all loads belonging to the northern part of the power grid are resilient against the perturbation and the network remains fully frequency synchronized. 
If any load belonging to the southern part of the power grid ($i=74$ and $i\geq77$) is perturbed, frequency synchronization across the power grid is lost. Moreover a shift in frequencies develops between the northern and southern part of the network, which is often accompanied by the full desynchronization of a generator ($i=76$) located near the boundary between the two parts. A visualization of the boundary of the perturbation is shown in Fig.\thinspace\ref{fig:VulnMaps}\thinspace (c).
A peculiar case is given by load $i=75$: irrespectively of being already in the southern, peninsular part, it is better connected to the northern part of the grid than to the southern part, therefore it is not affected by the increase of the demand and the system is able to compensate the perturbation of this node. In general, loads in the northern part of Italy have a larger connectivity than those in the south, thus local circuits are able to re-route the power flow and compensate for the single node demand. 
\\
In Fig.\thinspace\ref{fig:CMode_Comp_single} (g),\thinspace (h) the impact of applying control to the system is illustrated.
We can clearly observe the resilience of the nodes with the lower indices, which correspond to the northern part of the power grid, whose perturbation does not affect the frequency synchronization
of the grid. As before, we observe that \textit{difference control} is ineffective in restoring full frequency synchronization without the presence of additional links in the communication layer between the northern and southern part of the power grid (g). However, if additional links are present \textit{difference control} becomes very effective, while \textit{direct control} is unable to restore frequency synchronization within the network (h). \textit{Combined control} acts as effectively as the better of its two components.

\subsection{Perturbations targeting multiple generators simultaneouly}
In addition to targeting each individual generator in the network with different perturbations, we will now investigate the impact of targeting several generators simultaneously. For this purpose we will systematically increase the number $n$ of targeted generators.
Generators are perturbed successively from south to north along the Italian grid: the perturbation first affects nodes in the southern part of the network (characterized by higher node index), and generators with decreasing
node index are added successively, one by one, to the list of the perturbed nodes. Thus the perturbation propagates from the south to the north of Italy. 
\begin{figure}[ht]
		\includegraphics[width=1\linewidth]{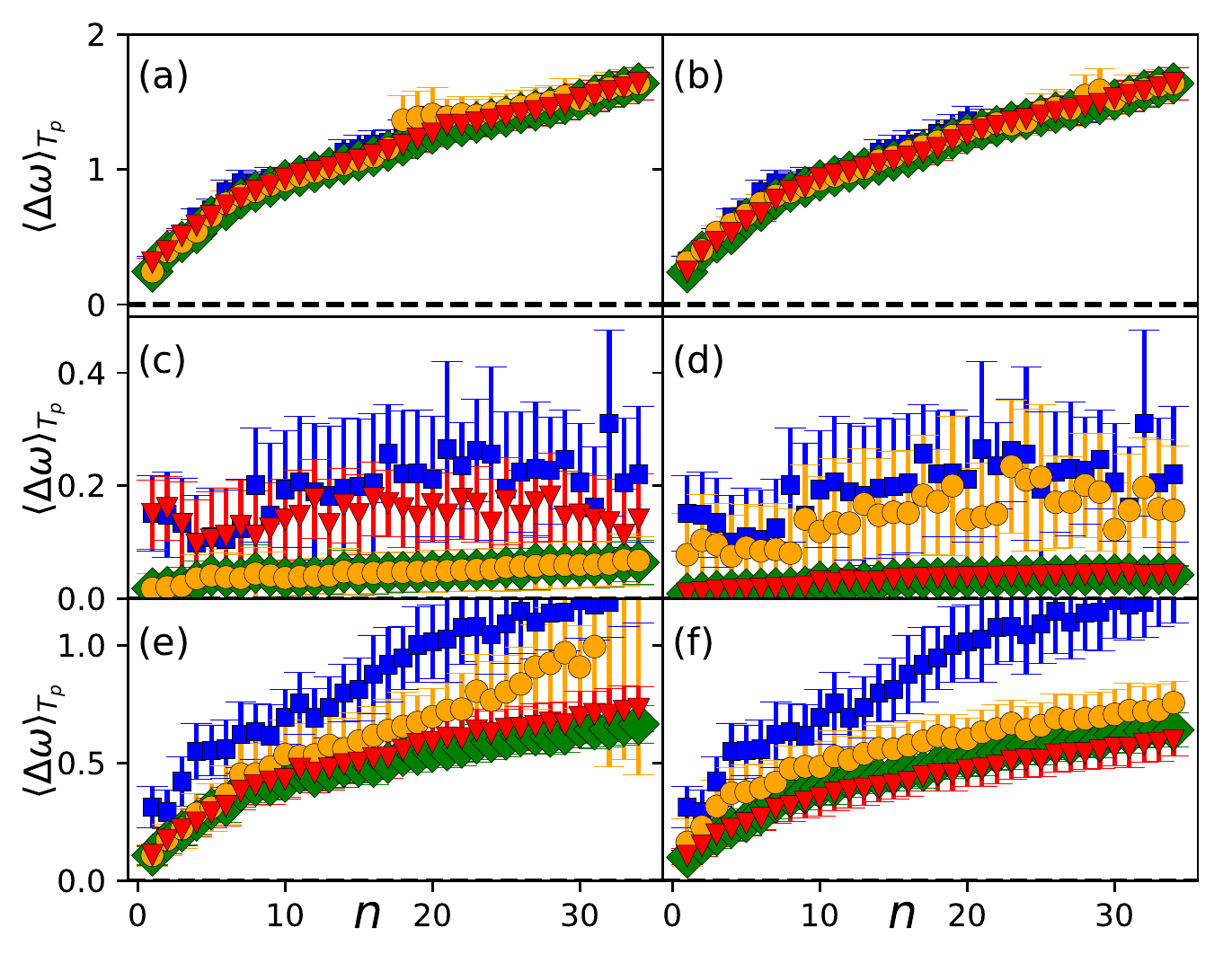}
	\caption{Control of frequency synchronization for different perturbations targeting multiple nodes: Mean standard deviation of frequency during the time of perturbation $\langle\Delta\omega\rangle_{T_p}$ vs the number $n$ of targeted nodes. 
		The symbols indicate different control schemes. Blue squares: no control. Red triangles: \textit{difference control}. Yellow circles: \textit{direct control}. Green diamonds: \textit{direct control}.
		Control strength $G=0.04$.
		The two columns show different control layer topologies, and the rows depict different types of perturbations:
		(a),(b): Disconnecting nodes using (a) $c_{ij}^{loc}$, (b) $c_{ij}^{ext}$.
		(c),(d): Intermittent noise using (c) $c_{ij}^{loc}$, (d) $c_{ij}^{ext}$.
		(e),(f): White noise using (e) $c_{ij}^{loc}$, (f) $c_{ij}^{ext}$.
		Parameters of the perturbations as in Fig.\thinspace\ref{fig:CMode_Comp_single}. Note the different vertical scales.
	}
	\label{fig:CMode_Comp_multi}
\end{figure}
Fig.\thinspace\ref{fig:CMode_Comp_multi} shows the impact of applying different types of perturbations (similar to Fig.\thinspace\ref{fig:CMode_Comp_single} (a)- (f)) to an increasing number of generators with and without control.
Disconnection of nodes quickly destroys synchronization in the network in absence of control due to the emergent strong fluctuations, as shown in Fig.\thinspace\ref{fig:CMode_Comp_multi}\thinspace (a), (b). In general, for $n\leq 13$ the perturbation is confined to the southern part of the grid, and results in a frequency shift between the northern and the southern part. For $14\leq n\leq 25$ the perturbation jumps over to the continental north-eastern part of the network, while for $n>25$ the perturbation finally reaches the northwestern part of the network. A special situation arises when generators $i=126, 121$ are disconnected, corresponding to $n=2$: The disconnection of node 121 cuts out a dead-tree composed of a chain of 6 nodes, thus isolating Calabria and Sicily, but resulting in an ineffective perturbation for the remaining grid. Note that since the averaged frequency deviation includes a growing number of disconnected nodes which oscillate at their natural frequency $\Omega_{gen}$, it increases strongly with $n$, and hence the overall improvement by the control is almost concealed on the vertical scale shown. Nevertheless the node-resolved data show that control is effective.

If intermittent noise is applied to multiple generators (Fig.\thinspace\ref{fig:CMode_Comp_multi}\thinspace (c), (d)), we generally observe a frequency shift between northern and southern parts with stronger fluctuations at the boundary of the two parts. When the perturbation ends, the shift may persist, or single-node desynchronization may occur. In particular if generators close to the boundary are perturbed (i.e., $i=76,71$), for $11\leq n\leq 15$ the fluctuations across the network persist after the end of the perturbation, and a single generator desynchronizes, $i=71$. The remaining network recovers frequency synchronization.

In case of Gaussian white noise (see Fig.\thinspace\ref{fig:CMode_Comp_multi}\thinspace (e), (f)), we observe that more and more nodes desynchronize from the rest of the network and revert to their natural frequencies, if the number of perturbed generators increases.
Moreover the loss of synchronization persists even after the end of the perturbation. For $n<15$ desynchronization is confined to the southern part of the Italian grid and a frequency shift between continental and peninsular parts can be observed. This means that we can safely disconnect some generators from the north in addition to all generators from the south, and still the desynchronization will not spread to the north. For $n\geq 15$, the perturbation spreads to the northern part of the grid. In particular, if the generator $i=58$ is affected by the noise, being located at the center of the northern part, synchronization is completely lost and the whole network is affected by the perturbation.
  
Now we discuss the different control strategies. We observe, in contrast to the case where single generators are targeted, that \textit{direct control} is not reliably the most effective control scheme in the absence of additional links between the generators in the control layer.
Especially when applying intermittent noise to most generators in the network (Fig.\thinspace\ref{fig:CMode_Comp_multi}\thinspace (c), (d)) we observe that the mean standard deviation is noticeable higher than for \textit{difference control} because neighboring generators are compensating each other. 
If one generator adopts a negative frequency, while the other has a positive frequency, \textit{direct control} will only ensure that their amplitudes are similar, as it only acts to restore the mean frequency in the neighborhood of the controlled node to the nominal frequency of the network.
The frequency deviation induced by the perturbation is the larger, the more nodes are affected simultaneously. Thus the reliability of \textit{direct control} deteriorates with the severity of the perturbation.
In the case of multiple generators connected in a chain, \textit{direct control} can even lead to runaway desynchronization, as the middle generator tries to compensate for its two neighbors which in turn act to compensate for the middle generator.
\\
\textit{Difference control} by itself is ineffective in restoring synchronization during the time of perturbation without additional links between the generators in the communication layer, but it becomes very effective at preserving frequency synchronization within the power grid, if all generators are connected in the communication layer.
\\
In the absence of additional links in the communication layer, \textit{combined control} is governed by the interplay of its two components. 
The \textit{difference control} is preventing the \textit{direct control} from stabilizing an equilibrium between the generators that is far from frequency synchronization, which greatly improves the effectiveness of merely \textit{direct control}.
In the presence of additional links between the generators in the communication layer the dynamics of \textit{direct control} is again mostly dominated by its \textit{difference control} component.

\subsection{Continuously increasing demand of all nodes simultaneously}

\begin{figure}[ht]
		\includegraphics[width=1\linewidth]{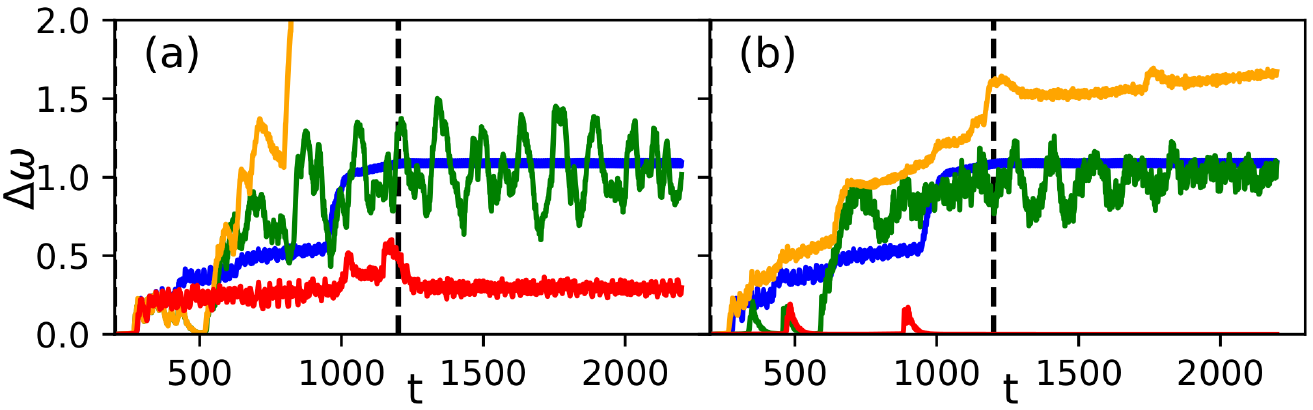}
	\caption{Control of frequency synchronization for continuously increasing demand of all nodes: Standard deviation of frequency $\Delta\omega$ vs time, when the demand of all loads in the network is simultaneously increased from $\Omega_{load}=-1$ to $\Omega_{pert}=-3$, starting at $t=200$. The final value is achieved at $t=1200$ marked by the dashed vertical line.
		Control strength $G=0.04$.
		The colours indicate different control schemes. Blue: no control. Red: \textit{difference control}, Yellow: \textit{direct control}. Green: \textit{direct control}.
(a) and (b) show different control layer topologies as in Fig.\thinspace\ref{fig:CMode_Comp_single}:
		(a) $c_{ij}^{loc}$,
		(b) $c_{ij}^{ext}$.
	}
	\label{fig:CMode_Comp_DemNet}
\end{figure}

Now we consider a perturbation of loads, where all loads increase simultaneously and continuously. This models the effect of external events, such as additional lighting needed during dusk or heat-induced usage of air-conditioning. As detailed in Sec.~\ref{perturbations}, 
each load increases from $\Omega_{load}=-1$ at a constant rate over 1000 time units until a final value of $\Omega_{pert}=-3$ is reached for all loads.
The overall impact of the perturbation is very strong, and the higher percentage of loads in the southern part of the grid with respect to the north plays a fundamental role in destabilizing the network
(see Fig.\thinspace\ref{fig:CMode_Comp_DemNet}\thinspace (a), (b)). 
As a result of this perturbation, generators at the boundary between north and south are the first to desynchronize, then fluctuations culminate in the desynchronization of multiple generators in the northern part. At this point the frequency shift between north and south disappears and the network shows a negative average mean frequency trying to compensate the desynchronized generators that oscillate at high positive average frequency.
\\
In Fig.\thinspace\ref{fig:CMode_Comp_DemNet} the effect of this perturbation on the frequency synchronization of the network and the ability of the different control strategies to preserve frequency synchronization are shown in more details.
We clearly observe that the only control scheme able to improve the uncontrolled network is \textit{difference control}.
\\ 
Without additional communication links between the generators, \textit{direct control} fails due to 
severe frequency deviations introduced by the perturbation on a small subset of generators in the northern part of the grid. Moreover the desynchronization of these generators ($i=22,\,57,\,58$) 
is reinforced by the control scheme, as neighboring generators try to compensate the frequencies of the others causing one of them to have increasingly negative frequencies, while the other shows increasingly positive frequencies. This can be mitigated by the introduction of additional communication links between the generators (b), but the control scheme still remains inefficient.
\\
\textit{Combined control} proves ineffective as well, regardless of the investigated control topology. \textit{Combined control} has been designed as a combination of \textit{difference control} and \textit{direct control}, but in this case the two components are competing against each other, causing the frequencies of the controlled generators to oscillate.
Since the mean frequency of the power grid is shifted away from the nominal frequency due to the perturbation, \textit{direct control} generally tries to increase the output of the generators to restore power balance, while the \textit{difference control} generally tries to match the decreased frequency of the loads to ensure frequency synchronization. This potential competition was non evident for the other perturbation schemes since in all previous cases there was no strong overall deviation from power balance across the network.

\subsection{Comparison of the control schemes}
\textit{Difference control} proves ineffective in the local control topology $c_{ij}^{loc}$, because it is only able to improve upon frequency synchronization locally. 
This is not due to a general malfunctioning of the control scheme, but rather to a limitation of the designed control in achieving full frequency synchronization: the control is always able to prevent lasting desynchronization of a node from the remaining grid, but not to prevent the desynchronization between the continental and the peninsular part of the grid.
This means that if a perturbation causes the northern and southern part of the network to lose synchronization, the control scheme only suppresses local fluctuations in the northern and southern part separately, but does not restore frequency synchronization between the two parts of the power grid: no generator is sufficiently connected to both parts simultaneously to make the control scheme effective. 
\\
This shortcoming is removed when we consider the extended control topology.
In particular, when additional communication links are introduced in $c_{ij}^{ext}$, all generators become well connected to both the northern and southern part of the grid, enabling the control to restore frequency synchronization across the whole grid. It is worth noticing that the extension of the average connectivity in the communication layer enables remote generators subject to perturbations to synchronize to unaffected generators, since the control scheme aims at synchronizing the frequency of the controlled node to its neighbors in the communication network, rather that synchronizing all nodes at a fixed frequency value. In this way the extended topology enables generators in the southern part of the grid, usually deeply affected by perturbations, to synchronize with generators in the northern part, that do not exhibit strong fluctuations.
%
%
\\
\\
\textit{Direct control} proves to be more effective in the local control topology $c_{ij}^{loc}$, where less communication links are present than in $c_{ij}^{ext}$. 
This is due to the basis mechanism underlying \textit{direct control}: it compensates the deviation of the mean frequency of all nodes connected to the controlled generator, thus causing the control to remain inactive when the mean frequency matches the nominal frequency of the power grid, while not all nodes are frequency synchronized.
On the other hand, this control scheme may become counterproductive if multiple generators are connected in a chain. If, for example, we consider the case of three generators connected in a chain, and the generator at one end adopts a negative frequency following a perturbation, it turns out that the middle generator adopts a positive frequency to restore the mean. The last generator in the chain then in turn adopts a negative frequency to compensate the middle one. 
The generator in the middle then begins compensating its two neighbors, which in turn try to individually compensate the middle generator.
This process leads to an unbounded increase in both negative and positive frequencies, which leads to quick desynchronization of the participating generators.
Adding further communication links in the control topology prevents this, but also renders the control scheme ineffective as multiple controlled generators compensate each other instead of restoring the nominal frequency within the power grid.
%
%
\\
\\
\textit{Combined control} in the local control topology $c_{ij}^{loc}$ is governed by its \textit{direct control} component. Since synchronization is lost across the grid, but not locally, \textit{difference control} is mostly inactive, while the \textit{direct control} part is responsible for restoring synchronization within the grid. The influence of the \textit{difference control} becomes visible in the cases where \textit{direct control} alone fails in the local topology, due to runaway interactions. These are successfully prevented by the \textit{difference control} component.
When additional communication links between the generators are present (i.e. $c_{ij}^{ext}$), the combined control is dominated by its \textit{difference control} component. In this case the \textit{direct control} is mostly inactive since the mean frequency across the two desynchronized regions of the network is equal to the nominal frequency. 
The drawback of applying both control schemes at the same time emerges when increasing demand of all loads in the network is implemented simultaneously. For this perturbation the two control schemes compete against each other, causing the frequency of the controlled node to oscillate, instead of being dominated by the more effective control scheme.

\section{Discussion}

The study of power grids is of great current interest in view of the impending termination of nuclear and fossil-fuel energy and replacement of old power plants with renewable energy sources \cite{UNF15}. Wind and photovoltaic power are the most promising technologies, but their integration into the grid poses a challenge \cite{JAC11,TUR99,UEC15}, in particular due to the fluctuation features of renewable power generators \cite{Anvari2016, Schmietendorf:EPJB:90} and the impact of energy trading on the network \cite{SCH18c}.

Besides these fluctuations it is also necessary to design modern power grids taking into account the topological change resulting from the increasing number of renewable sources which substitute fossil fuel power plants. This leads to the question of how the power grid should be structured and how the stability is modified due to decentralization. Recent work \cite{MEN14} found that the cost-minimizing creation of dead-end or dead-tree structures increases the vulnerability of the power grid to large perturbations, while a grid with decentralized power sources becomes more sensitive to dynamical perturbations, and simultaneously more robust to topological failures \cite{ROH12}. It is also possible that the addition of new power lines can decrease the stability of the power grid, which is known as the Braess's paradox \cite{WIT12}. Not negligible is also the role played by single critical nodes (dead-ends or frequency-deviating),
whose desynchronization is likely to result in a substantial blackout both in lossy \cite{HEL20} and lossless networks \cite{TAH19}.

These aspects present challenges to the stability of the power grid. During normal operation of the power grid, the network is in a synchronous state, in which all nodes run at the nominal frequency of the network ($50\,$Hz or $60\,$Hz). In case of excess demand, kinetic energy of synchronous generators is converted to electrical energy, thus decreasing the frequency. In order to cope with excess demand, usually primary control is achieved by the affected generator detecting this decrease and in turn increasing its power generation to restore the nominal frequency \cite{KUN94}, but the control of renewable power plants in this way is less feasible due to their reduced inertia \cite{ULB14,DOH10}.

Finally, even localized events can present a severe danger to the stability of the whole power grid, by causing a cascade of failures. Failure can occur due to multiple reasons, such as line overload, voltage collapse or desynchronization \cite{EWA78}. The dynamics of cascading failure, usually investigated in a monolayer power grid, shows the importance of considering transient dynamics 
of the order of few seconds \cite{SCH18z} since the distance of a line failure from the initial trigger and the time of the line failure are highly correlated. 

In this work we have conducted a systematic survey of controllability of the power grid against a variety of perturbations using the Italian grid as proof-of-principle. 
In particular we have presented a novel approach by considering the dynamics of a power grid in a two-layer network model, using a fully dynamical description for the communication layer. Previous investigations of multiple-layer power grids have been performed by taking into account only static nodes without dynamics, focusing on topological effects \cite{BUL10}. On the other hand, investigations of the dynamics of the (Italian) power grid are usually conducted only in a single layer \cite{serrani2004,fortuna2012,olmi2014,TAH19}, as well as the investigation of the dynamics of cascading failures \cite{SCH18z}.

Specifically, here we have modelled the Italian high voltage power grid as a dynamical two-layer network, where the dynamics of the power grid layer is described in terms of the second order Kuramoto model with inertia. On the other hand the second layer, which represents the communication network, models the dynamic control signal for each generator.
\\
We present the effects of a variety of different perturbations upon the frequency synchronization of the network. These perturbations model real-life threats to the stability of the power grid: Sudden failure of generators, increased demand by consumers, and one or more generators with stochastic power output fluctuations. The last of these perturbations is of particular interest, due to the challenge of replacing fossil fuel energy with renewable energy sources in the future.
To describe the fluctuating power output of renewable energy power plants both Gaussian white noise and more realistic intermittent noise have been used.
\\
In the communication layer we have assumed a selection of different control schemes (control functions $f_i^{diff}$, $f_i^{dir}$ and $f_i^{comb}$) and control topologies (adjacency matrices $c_{ij}^{loc}$ and $c_{ij}^{ext}$). All control schemes take advantage of the second layer by collecting information from adjacent nodes described by $c_{ij}$ to calculate the control signal. This can be done either in a local setting ($c_{ij}^{loc}$) where generators possess the same communication links as in the power grid layer, or in an extended control layer topology ($c_{ij}^{ext}$) where additional communication links between all generators are present. We have tested (i) a control scheme aimed at synchronizing the frequency of the controlled nodes with their neighbors (difference control $f^{diff}$), (ii) a control scheme aimed at restoring the original synchronization frequency in the neighborhood of the controlled node (direct control $f^{dir}$), and (iii) a mixed approach combining both ($f^{comb}$). The only control scheme being able to effectively counteract all of the perturbations is the difference control scheme $f^{diff}$ in the extended control topology, while the direct control has some advantages in the local control topology only.
\\
The investigation of the self-emerging control dynamics following perturbations has highlighted the role played by some specific nodes. In the Supplementary Material we have calculated the topological properties of the more vulnerable nodes in the southern region and at the boundary between north and south, as well as the less vulnerable nodes in the north-eastern part of the network.
The calculation of different topological measures shows that nodes in the power grid layer which are more affected by perturbations are not characterized, in general, by specific topological features. It turns out that the Italian power grid can be divided in two specific parts: the northern, continental part, with a higher average connectivity, which is more resilient to perturbations, and the southern, peninsular part, characterized by a low average connectivity. The elongated structure of the southern part makes it less robust to perturbations.
\\
Further work should incorporate time delay in the dynamics of the control grid, accounting for the delay in detecting the frequency of the generators and the delays during communication between them.
Furthermore the impact of an additional self-inhibiting term in the \textit{difference control} on its performance could be investigated.
Finally the perturbations in this work are each applied separately, but the interaction between them, especially of the stochastic fluctuations and the steady increase in demand, would be of interest.

\section{Methods}

\subsection{Power grid layer}
\begin{figure*}[ht]
		\includegraphics[width=0.32\linewidth]{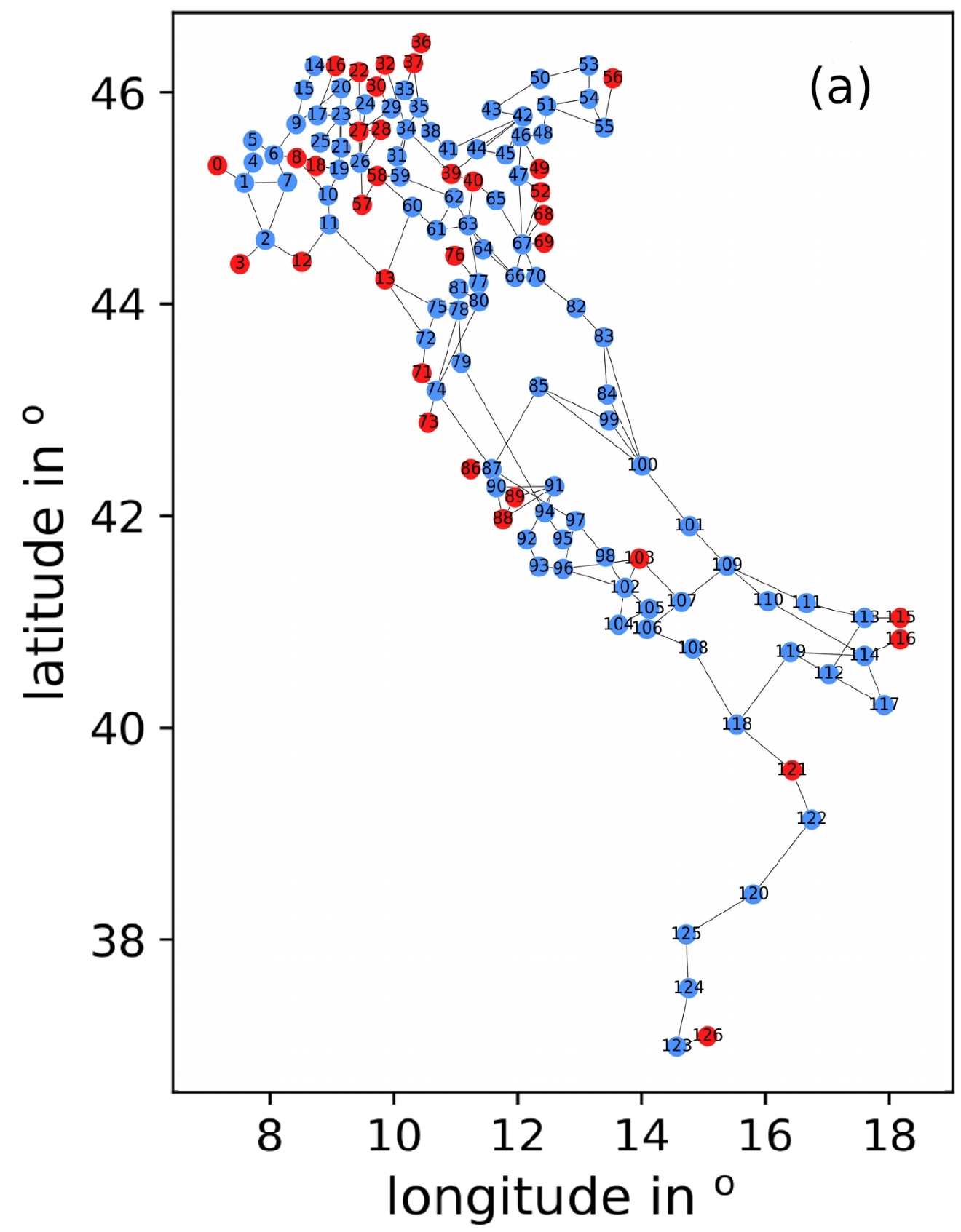}
		\includegraphics[width=0.32\linewidth]{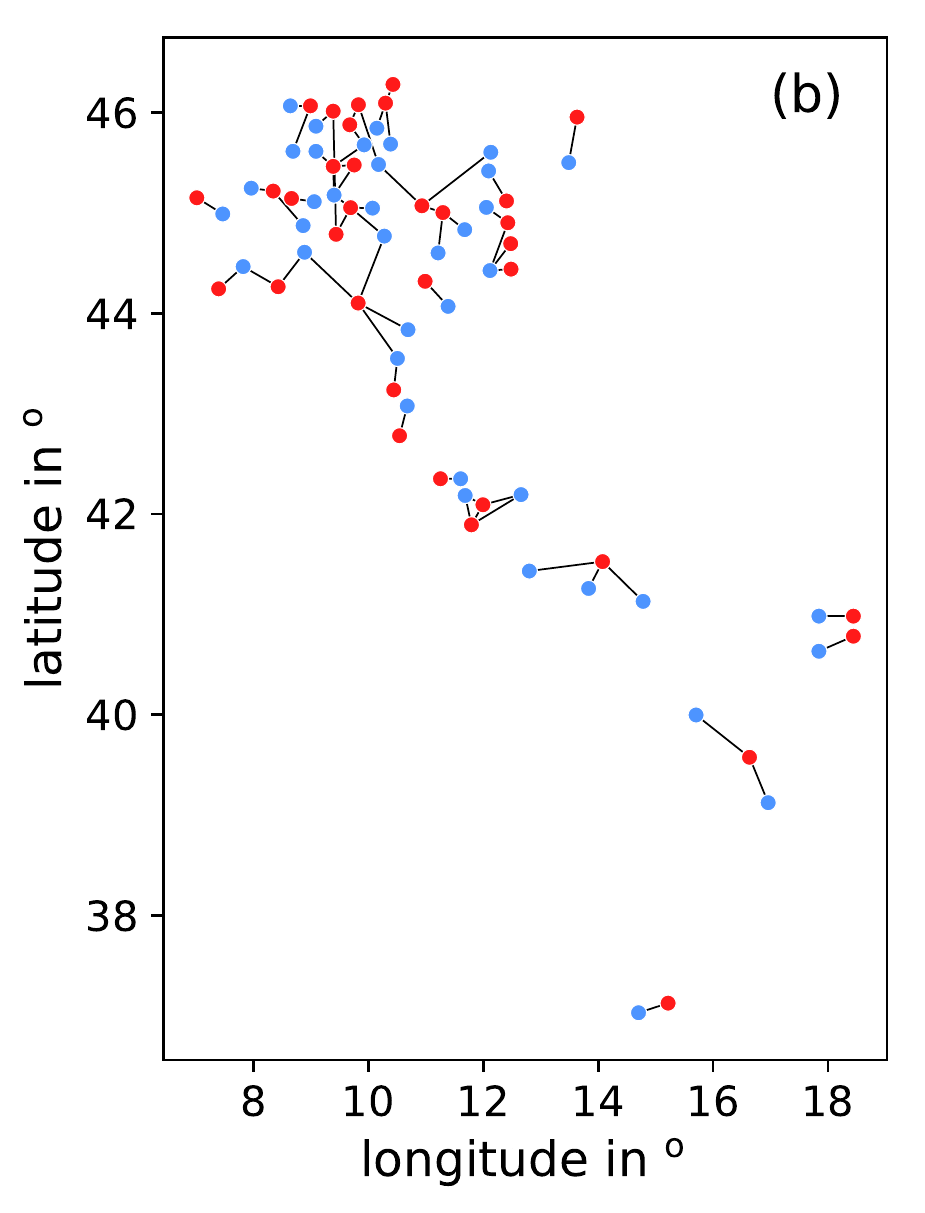}
		\includegraphics[width=0.32\linewidth]{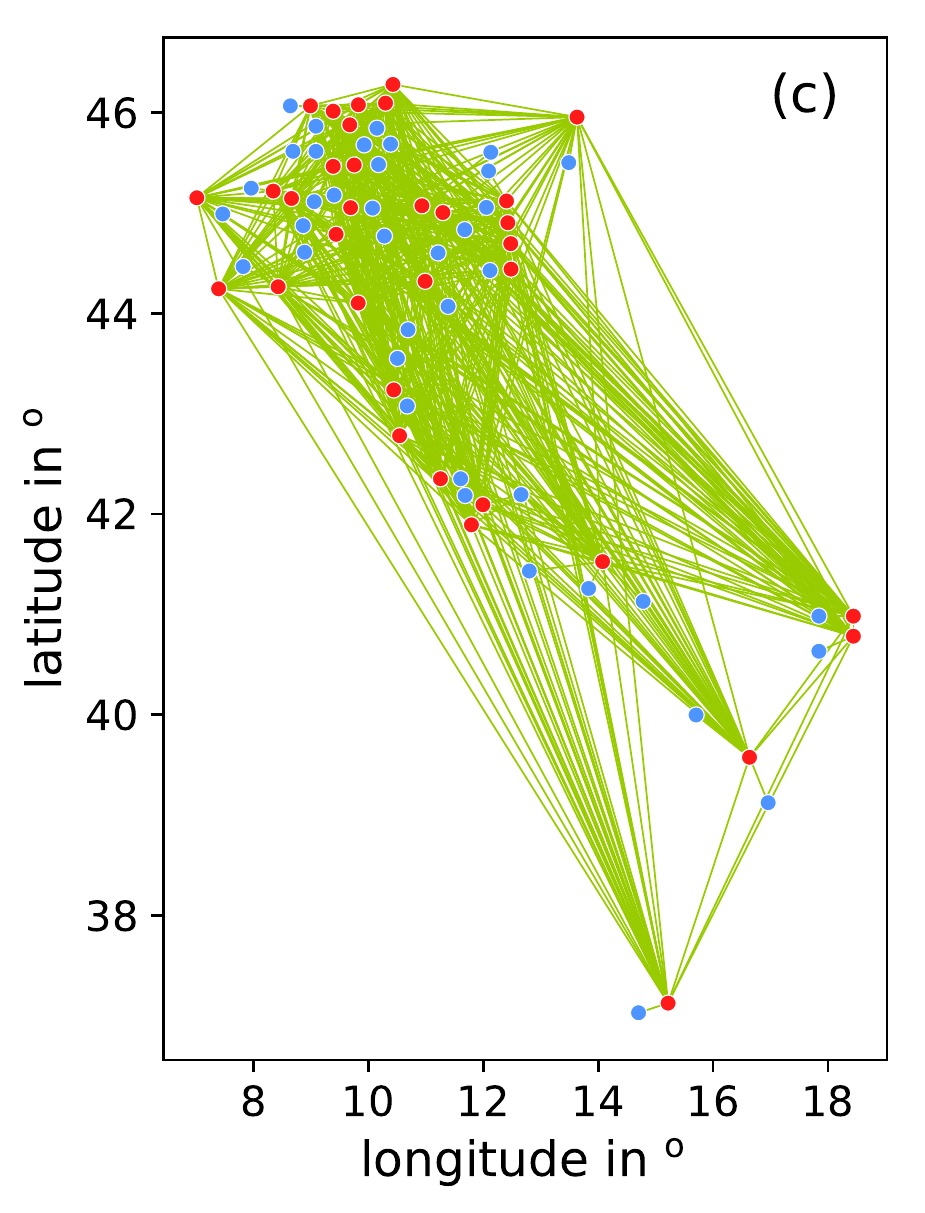}
	\caption{Visualisation of the topology of the individual layers of the two-layer power grid:
		(a) Topology $a_{ij}$ of the power grid layer based on the real Italian high voltage power grid;
		(b) control layer topology $c_{ij}^{loc}$ where the communication links of the generators are as in the power grid layer;
		(c) control layer topology $c_{ij}^{ext}$ where generators possess additional communication links to all other generators in the network (green). Red nodes denote generators, while blue nodes denote consumers. 
Position of nodes has been slightly modified to improve readability.
	}
	\label{fig:topologies}
\end{figure*}

The Kuramoto model with inertia describes the phase and frequency dynamics of $N$ coupled synchronous machines arranged in the controlled power grid layer, i.e., generators or consumers within the power grid, where mechanical and electrical phase and frequency are assumed to be identical:  
\begin{eqnarray}\nonumber
&& m\ddot{\vartheta_i}(t)= -\dot{\vartheta_i}(t) +\left(\Omega_i+P_i^c(t)\right) \\\label{dynamics-eq}
&+ &K\sum_{j}^N a_{ij}(t)\, \sin\left(\vartheta_j(t)-\vartheta_i(t)\right) \,,
\end{eqnarray}
with the phase $\vartheta_i$ and frequencies $\dot{\vartheta}_i$ of node $i=1,...,N$. Both dynamical variables $\vartheta_i$, $\dot{\vartheta}_i$ are defined relative to a frame 
rotating with the reference power line frequency (i.e., $50$ or $60 \, \text{Hz}$). The inherent frequency distribution is bimodal, where a positive natural frequency $\Omega_i$ of a node corresponds to the suitably normalized power supplied by a generator, while a negative natural frequency corresponds to the demand of a load.
The power balance requires that the power supplied by all generators in the network is exactly met by the combined demand of all loads: $\sum_i \Omega_i=0$.
The additional term $P_i^c$ denotes the control signal supplied by the communication layer, which serves as an offset to the power supplied by a controlled generator.
We assume homogeneously distributed transmission capacities $K$, while $m$ corresponds to the inertia.
The adjacency matrix $a_{ij}$ takes the value 1 if node $i$ has a transmission line connected to node $j$, 0 otherwise. In our numerical simulations we use the Italian high voltage power grid topology \cite{GEN19}, which consists of $N=127$ nodes, of which $34$ are generators and $93$ are loads. The matrix $a_{ij}$, which describes the topology, is unweighted and symmetrical (see Fig.\thinspace\ref{fig:topologies}\thinspace (a) for graph details). 

The Kuramoto model with inertia has been derived in \cite{Filatrella:EPJB:61} from the swing equation governing the rotor’s mechanical dynamics \cite{machowski2011}, by
assuming constant voltage amplitude and constant mechanical power $\Omega_i$. The former assumptions mean that the model does not describe voltage dynamics or the
interplay of amplitude and phase. A more realistic model would include the voltage dynamics, thus taking into account the machine's electrodynamical behavior, as done by \cite{schmietendorf2014}.
For simplicity also homogeneous transmission line capacities are assumed, and losses are neglected. A more realistic approach would
be to use individual transmission capacities to model different transmission line lengths, and to take into account sources of dissipation (e.g., Ohmic losses, and losses caused by damper windings) \cite{machowski2011}. However, the goal of the present paper is to gain insight into the principal behavior of large power grids depending on the network topology, and the interplay of the power grid layer with the communication layer, which is essential for the control mechanism. Therefore the assumptions we have made represent a good compromise when looking for universal properties: as always in modeling, the virtue of a simple model is that we can separate the influence of different
parameters and gain insight by studying a reduced model, while modelling all the details at the same time would not give insight, like a black-box brute-force numerical simulation. 

\subsection{Macroscopic indicators and parameter regime}
For our investigation we have chosen a regime of bistability in which both the fully frequency-synchronized state and a partially synchronized state are accessible. In this way 
it is possible to mimic the effect of a perturbation applied to the synchronized state; this displaces the system out of synchrony into an intermediate state where the operating conditions are not optimal for the functioning of power grids.

\begin{figure}[ht]
		\includegraphics[width=0.95\linewidth]{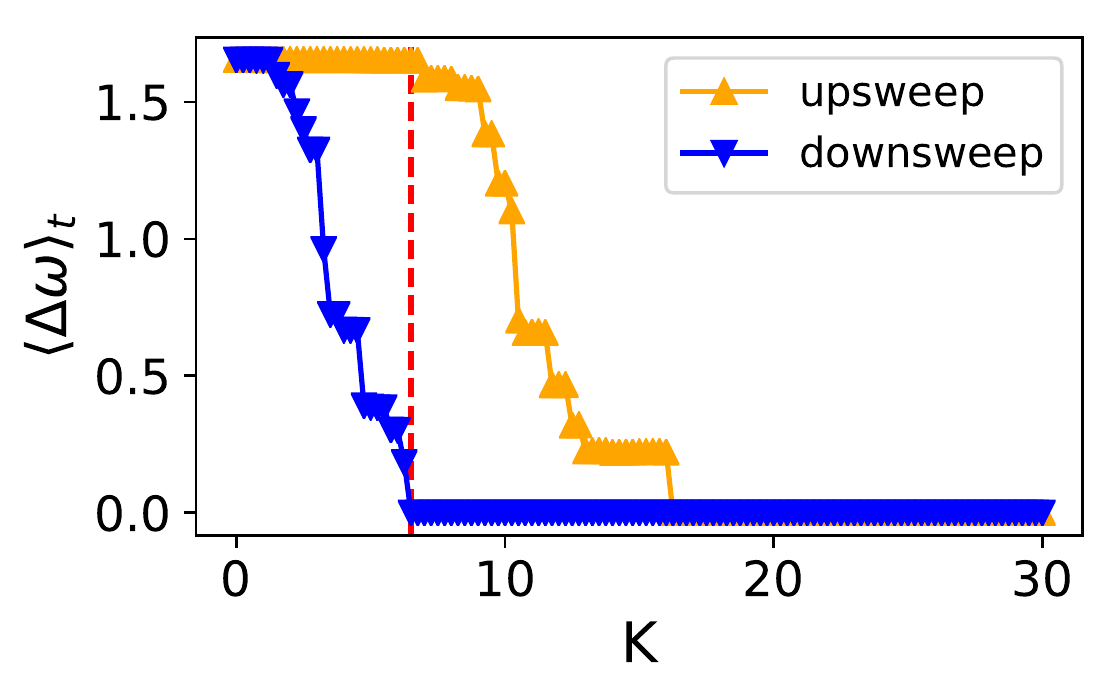}
		\includegraphics[width=0.95\linewidth]{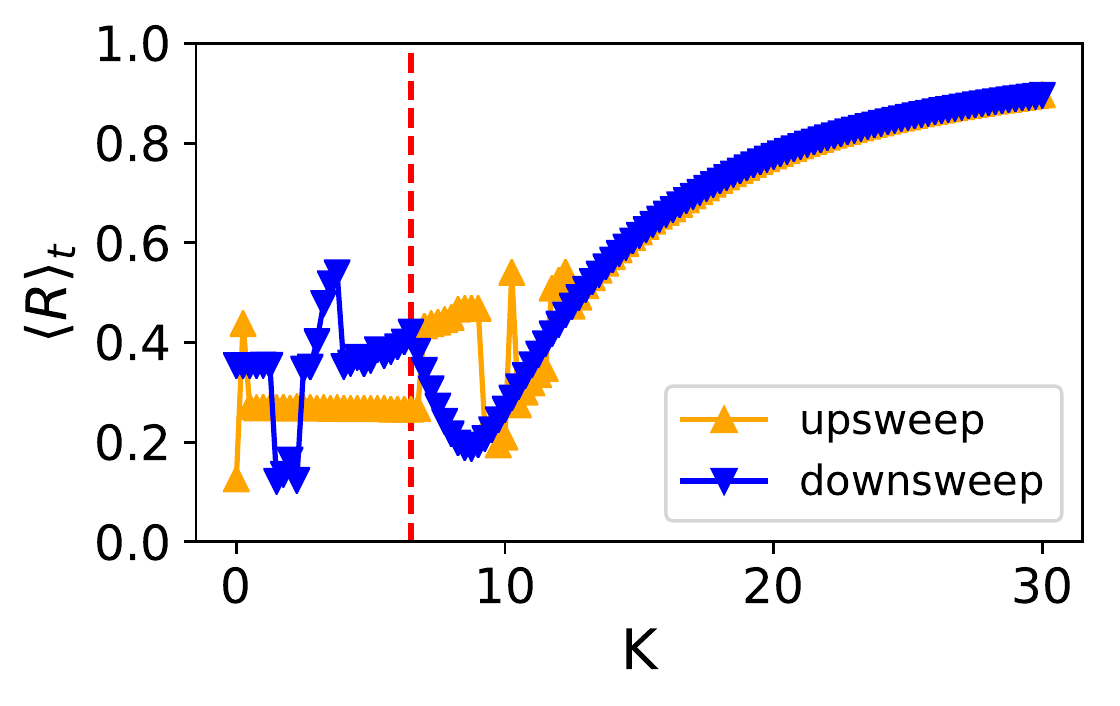}
	\caption{Mean synchronization of the power grid without control, averaged over $t_A=1000$ after a transient time of $t_R=10000$, vs coupling strength $K$. Upright orange triangles denote \textit{upsweep}, upside-down blue triangles denote \textit{downsweep}. The red vertical dashed line marks $K=6.5$.
		The synchronization is measured by:
		(a) mean standard deviation of frequency $\langle \Delta\omega\rangle_t$ (frequency synchronization),
		(b) mean Kuramoto order parameter $\langle R\rangle_t$ (phase synchronization). 	
		Parameters: $m=10$, bimodal frequency distribution with $\Omega_{load}=-1$ and $\Omega_{gen}=\frac{93}{34}$, integration time step $\Delta t=0.002$, adiabatic increase of $K$ with $\Delta K=0.25$.  
	}
	\label{fig:up/down-sweep}
\end{figure}

The system has been investigated in absence of control, by adiabatically varying the coupling strength $K$ with two different protocols \cite{TUM18,TUM19}. Namely, for the \textit{upsweep protocol}, the series of simulations is initialized for the decoupled system by considering random initial conditions both for phases and frequencies ($\vartheta_i(0)\in [-\pi, \pi)$, $\dot{\vartheta}_i(0)\in[-1,1)$ at $K=0$). Afterwards the coupling is increased in steps of $\Delta K$ until a maximum coupling strength is reached, where the system shows synchronized behaviour. For each investigated value of $K$, the system is initialized with the final conditions found for the previous coupling value. For the \textit{downsweep protocol}, starting from the maximum coupling strength achieved with the upsweep protocol, the coupling is reduced in steps of $\Delta K$ until the asynchronous state at $K=0$ is reached again. For each investigated coupling value the system evolves for a transient time $t_R$, until the network settles and it reaches a steady state. At this point, characteristic measures are calculated, averaging over a time $t_A$, in order to assess the level of synchronization of the state $\{\vartheta_i(t_R), \dot{\vartheta}_i(t_R)\}$.
In particular the time-averaged phase velocity profile $\langle\omega_i\rangle_t\equiv\langle\dot{\vartheta}_i\rangle_t$ provides information about the frequency synchronization of each node, while the standard deviation of the frequencies 
\begin{equation}\label{eq:average_stdev}
 \Delta\omega(t)=\sqrt{\frac{1}{N}\sum_{j}^{N}(\omega_j(t)-\overline{\omega}(t))^2}
\end{equation}
gives information about the deviation from complete frequency synchronization at the macroscopic level ($\bar{\omega}(t)$ represents the instantaneous ensemble-averaged grid frequency). 
In Fig.\thinspace\ref{fig:up/down-sweep}\thinspace (a) we depict the standard deviation of frequencies, averaged over a time interval $t_A$, $\langle\Delta\omega\rangle_t$ as a function of the coupling strength $K$. The difference between the results obtained for the upsweep protocol (orange triangles) and the downsweep one (blue triangles) highlights the hysteretic nature of the synchronization transition.
The phase ordering of the power grid is measured by the complex order parameter
\begin{equation}\label{EQ:order_parameter}
R(t) e^{i\Phi(t)} = \frac{1}{N} \sum\limits_{j=1}^N e^{i\vartheta_j},
\end{equation}
where the modulus $ R(t) \in \left [ 0 , 1 \right ] $ and the argument $ \Phi(t) $ indicate the degree of synchrony and mean phase angle, respectively. 
In the following we will denote $ R(t) $ as \textit{global order parameter}. In the continuum limit an asynchronous state is characterized by $R \approx 0$, while $R=1$ 
corresponds to full phase synchronization. Intermediate values of $R$ correspond to states with partial or cluster synchronization. The global order parameter, averaged in time, is shown as a function of $K$ in Fig.\thinspace\ref{fig:up/down-sweep}\thinspace (b). For small $K$ the state is asynchronous with $\langle R \rangle_t \simeq 1/\sqrt{N}$, then at $K\simeq 6.5$ for upsweep $\langle R\rangle_t$ exhibits an abrupt jump to a finite value, and then it decreases reaching a minimum at $K=9$. For larger $K$ the order parameter increases steadily with $K$ tending towards the fully phase synchronized regime, similarly to what was shown in \cite{olmi2014} for a different value of inertia.
In this article we have explored the dynamics of the system at $K=6.5$, where the system shows bistability between full frequency synchronization (i.e., it corresponds to the minimum coupling strength for which full frequency synchronization is still achievable) and partial synchronization ($\langle \Delta\omega\rangle_t\simeq 1.6$), which models the resulting state when the power grid is strongly perturbed.

\subsection{Communication layer}

\begin{figure}[ht]
		\includegraphics[width=0.95\linewidth]{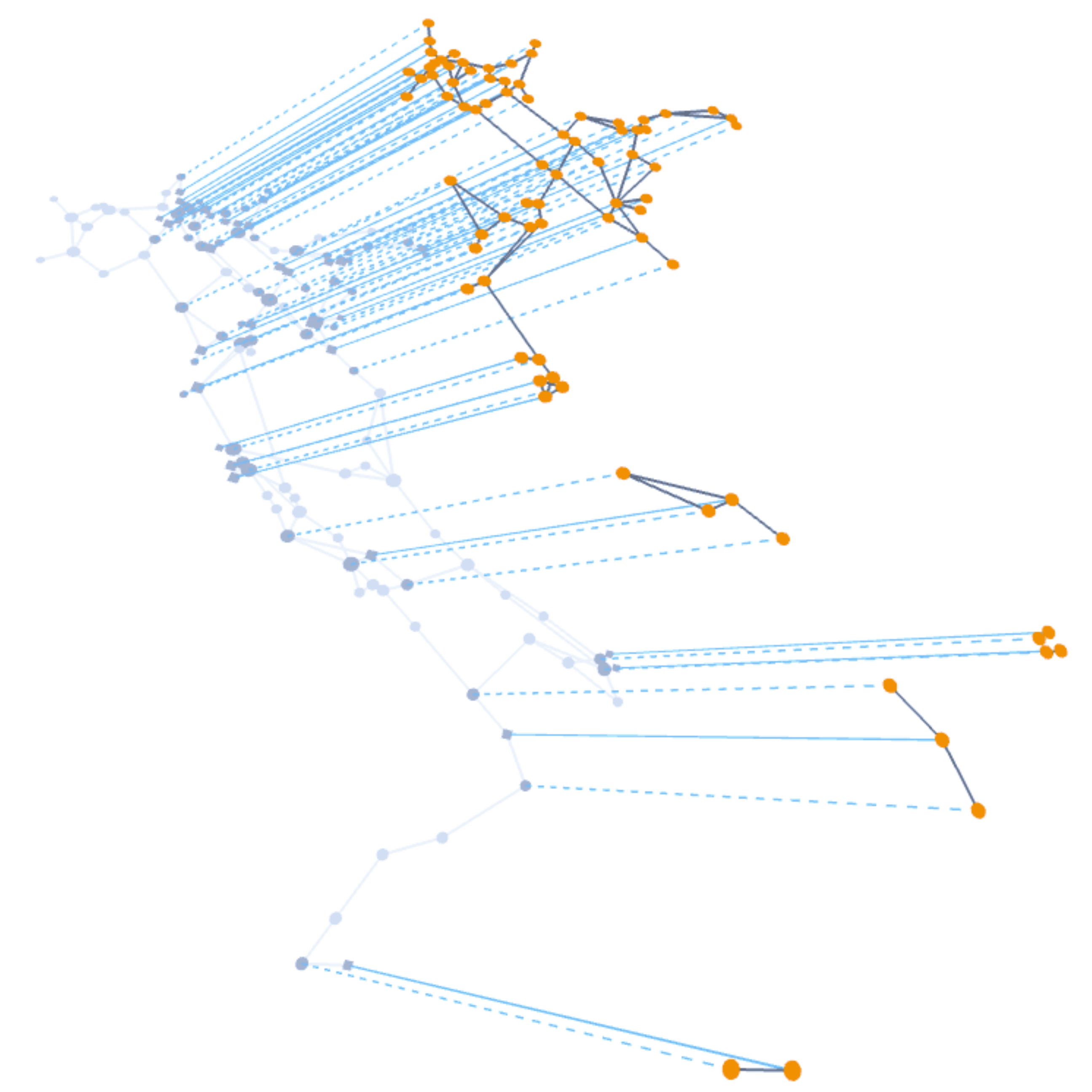}
		\caption{Graphical illustration of the two-layer network. Nodes of the
power grid (grey) are interconnected with the nodes of the communication network (orange). Straight blue lines indicate data flow from the generators to the communication layer, along with applied control input to the generators. Dashed blue lines represent data flow only from the loads to the communication layer.
}
		\label{fig_multilayer}
\end{figure}

The smart grid includes a communication infrastructure in all the stages of the power system, from transmission to users, allowing for the design of control strategies based on the information data flow. 
In real applications, we need to consider isolated elements, where synchronization needs to be assured such that, if the isolated nodes are reconnected to the main power system, failures can be avoided and the stability of the network is preserved. Therefore, the use of a communication layer of the network may improve the performance of the power system. We consider two layer topologies or infrastructures, the physical topology that describes the power system dynamics (as shown above in Fig.\thinspace\ref{fig:topologies}\thinspace (a)), and  the communication layer topology, which describes how data from each node is transmitted (see Fig.\thinspace\ref{fig:topologies}\thinspace (b), (c) for the topologies of the communication layer investigated here). Both infrastructures can be conceived as a multilayer network, as shown in Fig. \ref{fig_multilayer}, where for clarity we have chosen the communication layer topology presented in Fig.\thinspace\ref{fig:topologies}\thinspace (b).

To design a control strategy for synchronization, it is necessary to collect information from each generator and its neighbors. Phasor measurement units or sensors provide information, such that local controllers integrated with the generator nodes use the information to calculate a control signal $P^c_i\in \Re$. The loads are not controlled. The control signal can be interpreted as power injection for positive $P^c_i$ or power absorption for negative values of $P^c_i$, which is realized using storage devices (e.g., batteries) that can  absorb or inject power to the generator buses \cite{qian2010}. This real-world framework can be translated in terms of Eq. \ref{dynamics-eq} as injecting power in steady state operation.
 
Since the communication layer describes the exchange of information between the nodes about their current dynamic states, we consider a control strategy that depends on the information of the neighbors of each node, from which a control signal $P_i^c(t)$ for each controlled node is calculated dynamically. 
Neighbors can be related using the adjacency matrix $\mathcal{C}=\{c_{ij}\}$ of the communication layer. Our essential point is that we equip the communication layer also with a dynamics of its own. Therefore we propose to determine the control signal $P_i^c(t)$ by a first order differential equation depending on the frequencies $\dot{\vartheta_j}$ of neighboring nodes within the communication layer $c_{ij}$:
\begin{align} \label{control_general}
\dot{P_i^c}(t)=
G_i\,f_i\left(c_{ij},\{\dot{\vartheta_j}(t)\}\right) \,,
\end{align}
where $G_i$ is the control strength and $f_i$ represents the control function.
%

\paragraph{Control strength}
We assume that it is possible to control only the power output of generators in the network, thus $G_i$ is zero for all loads: 
\begin{align} \label{G_i_eq}
G_i= 
\begin{cases}
G \,\text{,} &i\in M_{gen}\\
0 \,\text{,} &\text{otherwise}
\end{cases}
\,
\end{align}
where $M_{gen}$ is the set of all generators in the network. Throughout this work we choose $G=0.04$.
%
\paragraph{Topology $c_{ij}$}
Two different topologies have been considered for the communication layer, namely $c_{ij}^{loc}$ and $c_{ij}^{ext}$.
In the local topology $c_{ij}^{loc}$ the connections between each generator and the other nodes in the communication layer correspond to the connections in the power grid layer (i.e., the communication layer network consists of a subnetwork of the power grid layer which contains all links of the generators in the power grid), except that each node has available also the information about itself, so that the diagonal elements of the adjacency matrix are nonzero.
The local topology is thus described by the adjacency matrix
\begin{align} \label{cij_loc_def}
c_{ij}^{loc}= 
\begin{cases}
1 \,\text{,} &i=j\\
a_{ij} \,\text{,} &\text{otherwise}
\end{cases}
\,.
\end{align}
As only generators receive a control signal $P_i^c$, all loads which are not connected to a generator can be disregarded in the communication layer, as illustrated in Fig.\thinspace\ref{fig:topologies}\thinspace (b).

For the extended control topology $c_{ij}^{ext}$ additional links between all generators are present (see Fig.\thinspace\ref{fig:topologies}\thinspace (c)). The globally coupled generators represent a subset of the communication layer (which additionally contains the neighboring loads); this provides an exchange of information between all generator control stations. The corresponding connectivity matrix is defined as
\begin{align} \label{cij_ext_def}
c_{ij}^{ext}= 
\begin{cases}
1 \,\text{,} &i\wedge j\in M_{gen}\\
c_{ij}^{loc} \,\text{,} &\text{otherwise}
\end{cases}
\,.
\end{align}
%
%
\paragraph{Control function $f_i$}
The dynamics of the control signal by Eq. \eqref{control_general} is governed by the control function $f_i\left(c_{ij},\{\dot{\vartheta_j}\}\right)$.
In this work we consider three different control functions: the first one is to apply a control signal that is proportional to the frequency difference between node $i$ and its neighbors \cite{Giraldo:CDC:2013}
\begin{align} \label{diffC-eq}
f_i^{diff}\left(c_{ij},\{\dot{\vartheta_j}(t)\}\right)=
\sum_j^N c_{ij}\left(\dot{\vartheta_j}(t)-\dot{\vartheta_i}(t)\right)
\,.
\end{align}
We refer to this control scheme as \textit{difference control}.
\\
The second approach is to apply a control signal that aims to restore power balance in the neighborhood of the controlled node:
\begin{align} \label{dirC-eq}
f_i^{dir}\left(c_{ij},\{\dot{\vartheta_j}(t)\}\right)=
-\frac{1}{N_i}\sum_j^N c_{ij}\,\dot{\vartheta_j}(t)
\,.
\end{align}
Here $N_i$ gives the number of direct neighbors of node $i$ in $c_{ij}$.
We will call this control scheme \textit{direct control}.
\\
The final control scheme is a combination of both difference and direct control:
\begin{align} \label{bothC-eq}
f_i^{comb}\left(c_{ij},\{\dot{\vartheta_j}(t)\}\right)=
\sum_j^N c_{ij}
\left(
a\left(\dot{\vartheta_j}(t)-\dot{\vartheta_i}(t)\right)
-b\frac{\dot{\vartheta_j}(t)}{N_i}
\right)
\,.
\end{align}
Here $a$ and $b$ are weight factors of the two components. 
In all further instances we will assume that $a=b=1$. 
We will refer to this control scheme as \textit{combined control}.

\subsection{Perturbations}
\label{perturbations}
We apply a selection of different perturbations to the network in order to test the efficiency of the control schemes. In the following the different kinds of perturbations are listed and discussed.
\paragraph{Disconnecting generators.}
The perturbed generator $i$ is disconnected from all its neighbors in both the power grid layer and the communication layer for a time interval $T_p$. This perturbation can be expressed in terms of the adjacency matrices of the power grid layer $a_{ij}$ in Eq.~(\ref{dynamics-eq}) and of the communication layer $c_{ij}$ in Eq.~(\ref{control_general}):
\begin{align} \label{PDisc-def}
\begin{cases}
a_{ij}\left( t\right) &= a_{ji}\left( t\right) =0\\
c_{ij}\left( t\right) &= c_{ji}\left( t\right) =0
\end{cases}
\,\,\,\, t\in T_P
\,.
\end{align}
This kind of perturbation represents a critical failure of a power plant.

\paragraph{Generators with fluctuating power output.}
Conventional power plants, whose power output is constant in time, and adjustable to the current demand, are progressively being replaced by stochastically fluctuating renewable energies like wind and solar. To represent the impact of renewable energy sources we investigate the impact of stochastic feed-in both with Gaussian characteristics and with more realistic properties (i.e., temporal correlations, realistic power spectrum and intermittent increment statistics).

Fluctuating power output can be modelled by modifying the natural frequency $\Omega_i$ of the targeted generator $i$ by adding a stochastic term $\Omega_N(t)$
\begin{align} \label{PNoise_def}
\Omega_i(t)=\Omega_{gen}+\Omega_N(t)
\,.
\end{align}

In particular, the stochastic term $\Omega_N(t)$ is chosen (i) as Gaussian white noise and (ii) as intermittent noise.
In case of Gaussian white noise 
\begin{align} \label{GWN_Def}
\Omega_N(t)=\sqrt{2D}\xi(t)
\, ,
\end{align}
where $\xi$ is a $\delta$-correlated Gaussian random variable, characterized by its noise intensity $D$.
\\
On the other hand, intermittent noise
\begin{equation}\label{noise_interm}
\Omega_N(t)=\mu x(t)
\end{equation}
is characterized by the penetration parameter $\mu$ and the intermittent noise series $x(t)$, whose generation is described in the following paragraph.
The main difference between intermittent noise and Gaussian white noise is that intermittent noise is highly correlated in time, and has fat tails at short correlation times, while Gaussian white noise does not show heavy tails nor any time correlation. This means that on short timescales extreme events are more likely to occur if intermittent noise is implemented.
\paragraph{Generation of intermittent noise.}
Due to atmospheric turbulence, wind power has specific turbulent characteristics \cite{MIL13, Anvari2016}, such as extreme events, time correlations, Kolmogorov power spectrum, and intermittent increment statistics. In particular the increment probability density functions of real wind power data significantly deviate from Gaussianity and its power spectrum displays $(5/3)$-decay with some discrepancy in the high frequency range. Based on this, we generate intermittent power time series $x(t)$ according to the synthetique feed-in noise generation detailed by Schmietendorf et al. in \cite{Schmietendorf:EPJB:90}. The first step in generating the intermittent noise time series is to consider the dynamics of the following Langevin-type system of equations:
\begin{align} \label{interm_noise_base}
\dot{z}(t)&=z(t)\left( g-\frac{z(t)}{z_0}\right) +\sqrt{Iz^2(t)}y(t)
\,,
\\
\dot{y}(t)&=-\gamma y(t) +\xi(t)
\,,
\end{align}
where $y(t)$ represents colored noise generated by an Ornstein–Uhlenbeck process \cite{GIL96,RIC88} with a correlation time $\tau_{OU}=1/\gamma$ and with a $\delta$-correlated Gaussian white noise term $\xi$.
The parameter $I$ controls the intermittency strength, while the other parameters $\gamma = 1.0$, $g = 0.5$ and $z_0 = 2.0$ are chosen as in \cite{Schmietendorf:EPJB:90}.
In a second step the time series $z(t)$ is transformed, so that its power spectrum resembles more closely the power spectrum of wind power plants.
To achieve this, the Fourier transform
$X(f)=FT\left[ z(t) \right] (f)$
is first divided by its amplitude spectrum.
This process eliminates the amplitude information of $X(f)$, but retains its phase information.
Subsequently a weight function $h(f)$ is used in order to make the spectrum of the series similar to the empirical data:
\begin{align} \label{IN_generation_2}
\hat{X}(f)=\frac{X(f)}{\left| X(f)\right|} h^{\frac{1}{2}}(f)
\,.
\end{align}
The power spectrum of $\hat{X}(f)$ is proportional to the weight function.
Finally $\hat{X}$ is transformed back into the time domain: $\tilde{x}(t)=FT^{-1}[\hat{X}(f)](t)$.
Due to the elimination of amplitude information the amplitude of $\tilde{x}$ is freely scalable. 
The standard deviation $\sigma_{\tilde{x}}$ of the aggregated distribution of $\tilde{x}$ is rescaled to any desired $\sigma_{{x}}$:
\begin{align} \label{IN_final}
{x}(t)=
\frac{\sigma_{{x}}}{\sigma_{\tilde{x}}}\tilde{x}(t)
\,.
\end{align}
Since $\langle x(t)\rangle = 0$, $\langle\Omega_i(t)\rangle=\Omega_{gen}$ and power balance is maintained on long-time average.

Further restrictions are introduced to make this perturbation more realistic.
A lower boundary for ${x}$ is introduced, so that a generator cannot operate as a load in the network due to the influence of noise: all values $x<-1$ are truncated to $x=-1$.
Furthermore a power feed-in cut-off is assumed, which means that generators have a maximum power output they can supply:
all values $x>1$ are truncated to $x=1$.
This additionally truncates some of the extreme events in the strongly intermittent power time series, while the mean and standard deviation are nonetheless almost unaffected by this.
With these constraints for ${x}$, the penetration parameter $\mu$ is chosen equal to the natural frequency $\Omega_i$ of the affected generator.
\\
Throughout this work we fix $I=2$,  $h(f)=f^{-\frac{5}{3}}$ and $\sigma_x=\frac{1}{3}$ for intermittent noise. 


\paragraph{Increasing demand of loads.}
First, when targeting a single load $i$, it is subjected to an instantaneous increase in demand. This is realized by increasing its natural frequency to $|\Omega_{pert}|>|\Omega_{load}|$ for the duration of the perturbation $T_p$, where $\Omega_{load}<0$ gives the natural frequency of the load during normal operation of the power grid. Then the perturbation is
\begin{align} \label{PDemandS-def}
\Omega_i(t)=
\begin{cases}
\Omega_{pert} \,\text{,} &t\in T_P\\
\Omega_{load} \,\text{,} &\text{otherwise}
\end{cases}
\,.
\end{align}
Furthermore, we investigate the effect of steadily increasing the demand of all loads simultaneously. 
For all loads experiencing the perturbation, the evolution of the natural frequency is described by
\begin{widetext}
 \begin{align} \label{PDemandNetwork-def}
\Omega_i(t)=
\begin{cases}
\Omega_{load} \,\text{,} &t<t_{start}\\
\Omega_{load}+
\left(\Omega_{pert}-\Omega_{load}\right)
\frac{t-t_{start}}{t_{end}-t_{start}}   
\,\text{,} &t_{start}\leq t\leq t_{end}\\
\Omega_{pert} \,\text{,} &t<t_{end}
\end{cases}
\,,
\end{align}
\end{widetext}
where $t_{start}$ denotes the time when the demand of all loads begins to increase and $t_{end}$ denotes the time when the final value of demand is reached.
\\
This perturbation is suited to represent a change in consumer behavior due to external circumstances such as, e.g., dawn and dusk, or hot and cold weather.

\subsection{Numerical integration}
To integrate Eqs. \eqref{dynamics-eq}, \eqref{control_general} and solve the dynamics of the system, we use the fourth order \textit{Runge-Kutta}-algorithm \cite{SAN00} with an integration time step of 
$\Delta t=0.002$. 
When applying noise to the system the integration scheme is changed according to the applied noise scheme. 
For intermittent noise Eq.~\eqref{IN_final}, the noise series $x_i(t)$ affecting a single node are pre-calculated and used in Eq.~\eqref{dynamics-eq} as a modification of the generator's constant power input, see Eq. \eqref{noise_interm}. The dynamics is solved by simply applying a fourth order \textit{Runge-Kutta} integration scheme with a time step of $\Delta t=0.001$.
When applying Gaussian white noise to the generators, we cannot use the standard Runge-Kutta methods, therefore we employ a method similar to the second–order Runge–Kutta, which has been developed in \cite{SAN00} on the basis of the \textit{Heun} algorithm. For the latter algorithm we use an integration time step of $\Delta t=0.001$.  

\section*{Data Availability Statement}
The data that supports the findings of this study are available within the article.

\section*{Acknowledgements}

The authors acknowledge Enrico Steinfeld and Katrin Schmietendorf for useful discussions. 
Funded by the Deutsche Forschungsgemeinschaft (DFG, German Research Foundation) - Projektnummer 163436311 - SFB 910.

\section*{Author contributions statement}

S.O. and E.S. conceived and designed the research. C.T. performed the simulations. All authors 
contributed to discussing the results and writing the manuscript.

\section*{Competing interests}
The authors declare no competing financial interests.

\bibliography{apsrev4-1}

%





\end{document}